\documentclass[useAMS,usenatbib]{mn2e}
\usepackage{amsmath}
\usepackage{rotating}
\usepackage{graphicx}

\newcommand\lsim{\mathrel{\rlap{\lower4pt\hbox{\hskip1pt$\sim$}}
        \raise1pt\hbox{$<$}}}
\newcommand\gsim{\mathrel{\rlap{\lower4pt\hbox{\hskip1pt$\sim$}}
        \raise1pt\hbox{$>$}}}
\newcommand{\apj}{ApJ}

\newcommand{\apjl}{ApJL}
\newcommand{\mnras}{MNRAS}

\newcommand{\apjs}{ApJS}

\newcommand{\pasj}{PASJ}

\newcommand{\prd}{Phys. Rev. D.}

\title[Gas Disk Response to Black Hole Recoil and Mass Loss]{Hydrodynamical Response of a Circumbinary Gas Disk \\ to Black Hole Recoil and Mass Loss} 
\author[Corrales et al.]{Lia R. Corrales$^1$, Zolt\'an Haiman$^1$, Andrew MacFadyen$^2$ \\
$^1$Department of Astronomy, Columbia University, 550 West 120th Street, New York, NY 10027 \\
$^2$Center for Cosmology and Particle Physics, Physics Department, New York University, New York, NY 10003}

\begin{document}

\date{\today}
\maketitle


\begin{abstract}
  Finding electromagnetic (EM) counterparts of future gravitational
  wave (GW) sources would bring rich scientific benefits.  A promising
  possibility, in the case of the coalescence of a super-massive black
  hole binary (SMBHB), is that prompt emission from merger-induced
  disturbances in a supersonic circumbinary disk may be detectable.
  We follow the post--merger evolution of a thin, zero-viscosity
  circumbinary gas disk with two-dimensional simulations, using the
  hydrodynamic code FLASH.  We analyze perturbations arising from the
  530 ${\rm km~s^{-1}}$ recoil of a $10^6 M_\odot$ binary, oriented in
  the plane of the disk, assuming either an adiabatic or a
  pseudo--isothermal equation of state for the gas.  We find that a
  single-armed spiral shock wave forms and propagates outward,
  sweeping up $\sim 20\%$ of the mass of the disk. The morphology and
  evolution of the perturbations agrees well with those of caustics
  predicted to occur in a collisionless disk. Assuming that the disk
  radiates nearly instantaneously to maintain a constant temperature,
  we estimate the amount of dissipation and corresponding post-merger
  light--curve.  The luminosity rises steadily on the time--scale of
  months, and reaches few $\times 10^{43}$ erg/s, corresponding to
  $\approx 10\%$ of the Eddington luminosity of the central SMBHB.  We
  also analyze the case in which gravitational wave emission results
  in a $5\%$ mass loss in the merger remnant.  The mass-loss reduces
  the shock overdensities and the overall luminosity of the disk by
  $\approx 15-20\%$, without any other major effects on the spiral
  shock pattern.
\end{abstract}

\begin{keywords}
black hole physics -- galaxies: nuclei -- gravitational waves
\end{keywords}


\section{Introduction}

The gravitational waves (GWs) produced during the late stages of the
merger between super-massive black holes (SMBHs), with masses of $\sim
(10^4$--$10^7)\,{\rm M_\odot}/(1+z)$, out to redshifts beyond
$z\approx 10$, are expected to be detectable in the next decade by the
{\it Laser Interferometric Space Antenna} ({\it LISA}) satellite.
While the GW signatures themselves will be a rich source of
information, identifying the electromagnetic (EM) counterpart of the
LISA source would open up a whole new range of scientific
opportunities, from black hole astrophysics to fundamental aspects of
gravitational physics and cosmology
\citep{koc07,HKM09a,Phinney,Bloom}.

Whether the EM counterparts of the SMBHBs detected by {\it LISA} can
be uniquely identified depends on the accuracy of localization by {\it
  LISA}, and on the nature of the EM emission.  For the typical SMBHB
detected by {\it LISA} at $z\approx 1-2$, the sky position and the
redshift will be known to the accuracy of $\delta(\Delta\Omega)\sim$
few $\times$ 0.1 square degrees, and $\delta z\approx$ few $\times$
0.01, respectively (the latter limited by weak lensing errors; e.g.
\citealt{hh05,koc06}).  Within this three--dimensional error volume,
there will be of order $\sim 100$ candidate galaxies, at the optical
magnitude limit expected to host such SMBHBs \citep{koc08}.  If the
coalescing SMBHs themselves produce bright emission comparable to
luminous quasars, or are associated with some other, similarly rare
subset comprising $\lsim 1\%$ of all galaxies (such as ultra--luminous
infrared galaxies), then a unique counterpart may be identified among
these candidates \citep{koc06}.  However, having a prediction for the
spectrum and the light--curve of a coalescing SMBHB -- and therefore
knowing what characteristic signatures to look for -- would make such
identifications both more likely and more reliable.

A promising possibility is that the coalescing SMBHB produces a {\em
  variable or transient} EM signal, which can be uncovered by suitably
designed EM observations, either concurrently with or following the
{\it LISA} detection \citep{koc08,lh08,HKM09a}.  The dense nuclear gas
around the BH binary is expected to cool rapidly, and settle into a
rotationally supported, thin circumbinary disk
\citep[e.g.][]{barnes,elcm05}.  The evolution of a SMBHB embedded in
such a disk has been studied in various idealized configurations
\citep[e.g.][]{an02,liu03,mp05,dot06,mm08,hayasaki09,cuadra09,HKM09b}.
Generically, at small orbital separations when the binary is
detectable by {\it LISA}, the orbit decays rapidly due to GW emission.
If the disk is thin, the binary torques create a central cavity,
nearly devoid of gas, within a region about twice the orbital
separation \citep[e.g.][]{al94}.  Whether the decaying SMBHBs produce
bright emission during this stage is not well understood.  If the
central cavity were truly empty, no gas would reach the SMBHBs, and
any emission produced farther out in the disk would likely be weak. On
the other hand, numerical simulations suggest residual gas inflow into
the cavity \citep{al96,mm08,hayasaki07,hayasaki08,cuadra09}, which may
plausibly accrete onto the BHs, producing non--negligible EM emission.

A different possibility, and the focus of the present paper, is that
variable EM signatures are produced in the gas disk {\it promptly
  after} the coalescence of the SMBHB.  The burst of GWs emitted
during the last stages of the coalescence results in a corresponding,
nearly instantaneous reduction in the binary's rest mass.
Furthermore, when compact objects coalesce asymmetrically, the linear
momentum carried away by the GWs imparts a ``kick'' to the center of
mass of the system \citep{Bekenstein, Fitchett}.  Recent
break--through in numerical relativity has allowed accurate
calculations of these effects, showing that the mass loss is typically
several percent \citep[e.g.][and references therein]{tm08}, while kick
velocities are typically several hundred ${\rm km~s^{-1}}$, but can be
as high as 4,000 ${\rm km~s^{-1}}$
\citep{baker06,baker07,baker08,campanelli07a,campanelli07b,gonzalez07a,gonzalez07b,
  herr07a,herr07b,herr07c,koppitz07}.  The circumbinary gas will
respond promptly (on the local orbital timescale) to such dynamical
disturbances.  Importantly, the gaseous disk outside the inner cavity
is expected to cool efficiently and become geometrically thin,
implying that the orbital motion of the gas is supersonic.  This gas
is therefore susceptible to prompt shocks, which could, in principle,
produce a detectable transient EM signature \citep{mp05}.

\citet{Lippai} considered the motion of collisionless test particles
in such a disturbed disk.  As long as the particles remain bound to
the central SMBHB, they follow elliptical Kepler orbits (in the
inertial frame of the SMBHB).  These orbits cross, and produce a
characteristic, outward--propagating spiral caustic.  While these
results were obtained in a pressureless ``dark matter'' disk, they
suggest that shocks, with a similar pattern, will indeed arise in the
gas, on times--scales of $\sim$weeks to $\sim$months after the
coalescence.  Similar conclusions were reached by \citet{Shields} and
by \citet{Schnittman}, using N-body simulations and semi--analytical
arguments, respectively, to show that a bright, post--merger ``flare''
may occur.  Both of these studies focused on the evolution of disks
around more massive BHs on longer ($\sim 10^4$yr) time--scales, and
proposed detecting the flare by monitoring a population of active
galactic nuclei (AGN).  By comparison, our analysis here focuses on
the disks around lower--mass BHs and on the shorter time--scales of
several weeks to a year, which are relevant for follow--up
observations triggered by {\it LISA} detections.

Motivated by the above, in the present paper, we follow up on earlier
work, and compute the response of the gas disk including the effects
of gas pressure using hydrodynamical simulations.  Our main goals are
(i) to assess shock formation in hydrodynamical disks, and (ii) to
estimate the amount of dissipation and the resulting light--curve
produced by the disk.  In particular, for the latter, we will use
simulations with a pseudo--isothermal equation of state. This
corresponds implicitly to strong dissipation, and should represent an
approximate upper limit on the disk luminosity.  Our computations are
performed with the publicly available code FLASH \citep{Fryxell}.

Two other, independent studies have recently used 3--dimensional
simulations to address the response of gas disks to mass--loss
\citep{ONeill} and to both mass--loss and kicks \citep{Megevand}.  The
main difference between our analysis and these previous works is our
inclusion of runs with a pseudo--isothermal equation of state, which
modifies, qualitatively, the expected EM signature.  In particular,
\citet{ONeill} and \citet{Megevand} both run simulations with an
adiabatic equation of state, and find that in most cases, the gas disk
typically {\em dims} following the merger. We observe a similar effect
in our adiabatic runs, and attribute it to an overall dilution of the
gas density.  However, the gas develops significantly larger density
contrasts in our pseudo--isothermal runs, and we argue that this can
result in a significant {\it brightening} of the system.  Another
important difference between our study and those of \citet{ONeill} and
\citet{Megevand} is that we perform two--dimensional simulations,
whereas \citet{ONeill} and \citet{Megevand} both utilized 3D
simulations.  While we have sacrificed resolving the vertical disk
structure, this allows us to simulate a disk that extends to much
larger radii ($10^4$ Schwarzschild radii), and follow the entire disk
for a much longer period ($\sim1$ yr for a $10^6~{\rm M_\odot}$
binary), in order to cover the regime relevant for follow-up
observations of {\it LISA} events.  Given the expected size of the
inner cavity in the disk, $\sim 100 R_s$, studying these outer regions
of the disk, and following the disk evolution on the correspondingly
longer time--scale, is especially important. Further differences
between our study and those of \citet{ONeill} and \citet{Megevand}
will be discussed in \S~\ref{sec:Others} below.

The rest of this paper is organized as follows.
In \S~\ref{sec:ProblemSetup}, we describe our computational setup,
including details of the simulations, and initial conditions.
In \S~\ref{sec:Results}, we present and discuss our main results.
These include the impact of the kick (\S~\ref{sec:PostMerger}) and the
mass--loss (as well as both effects in combination;
\S~\ref{sec:MassLoss}), using constant surface density disks.  In
\S~\ref{sec:Observational}, we discuss our estimate of the
post--merger light curves.  In \S~\ref{sec:Alphadisks}, we repeat our
calculations at higher resolution and with more realistic initial gas
density and temperature profiles (adapted from a standard
$\alpha$-disk).  In \S~\ref{sec:Numerical}, we discuss possible
numerical issues, and in \S~\ref{sec:Others}, we compare our results
to previous work.
Finally, in \S~\ref{sec:Conclude}, we summarize our main conclusions
and their implications, and outline natural future extensions of this
study.
%


\section{Initial Conditions and Simulation Details}
\label{sec:ProblemSetup}

Our calculations were performed with version 2.5 of the publicly
available Advanced Simulation and Computing (ASC) Flash code,
developed at the University of Chicago.  This code uses the split
piece-wise parabolic method (PPM), making it well-suited for modeling
shocks \citep{Fryxell}.  We use a uniform--mesh grid in polar
coordinates to simulate a two--dimensional vertically averaged disk.
Our runs cover a physical duration much shorter than the viscous
time--scale, and we therefore do not include any physical viscosity.
The central black hole is chosen have a mass of $M=10^6 {\rm
  M_\odot}$, which is approximately the most favorable value for
detection by {\it LISA} (although our results can be re--scaled to
apply to BHs with different masses; see \S~\ref{sec:Numerical}).  The
inner edge of the grid is located $100 R_s$ (Schwarzschild radii), the
radius at which the central gap opened by a binary black hole is
expected to freeze \citep{mp05,HKM09b}.

A high--resolution mesh is used for radii between $100 < R/R_s < 10^4$
for all simulations, except the runs \textsl{Kad1-5} (see
Table~\ref{tab:CDsims} below), which use a high--resolution mesh
between $ 100 < R / R_s < 1.5\times10^4$.  A lower--resolution mesh
extends to $3.1 \times 10^{5} R_s$.  We tested the effect of boundary
conditions by simulating a disk in Keplerian motion without the
influence of mass loss or kick.  Within the 486 days we simulate after
the merger, numerical fluctuations contributing to the density and
luminosity of the disk were confined within the $100 < R/R_s < 10^3$
region.  The amplitude of the numerical density perturbations was well
below the over--densities produced by shocks in the disturbed disks,
so in our analysis of the densities and morphologies of the shocks, we
include the entire high--resolution mesh region, down to $100 R_s$.
However, the contribution of this low--amplitude noise to the
luminosity was still significant.  As a consequence, our
post--processing analysis related to the disk luminosity is restricted
to radii between $10^3 < R/R_s < 10^4$; corresponding to $30~{\rm
  days}\lsim t \lsim 1 {\rm year}$ (see discussion of this issue in
\S~\ref{sec:Numerical} below).

In our fiducial run (see ``\textsl{Kiso3}'' in Table~\ref{tab:CDsims}
below), the high--resolution grid contains zones with sizes $\Delta r
= 20 R_s$ and $\Delta\theta = 0.01$ radians, which is the effective
spatial resolution of this run.  This run had a corresponding
numerical time--resolution, in physical units, of $\Delta t \approx
35$ seconds. However, to save disk space, FLASH outputs were made only
at every $\approx 10^4$th time-step, allowing us to sample the results
on time--scales of $\approx4$ days.  In physical units, the total
simulated time interval is 486 days (or $\sim 1.4\times10^7M$, in
gravitational units with $G=c=1$).  For comparison, the orbital time
at the inner and outer edge of our disk is $\sim 4$ hours and 0.5
years, respectively, so that our simulations cover approximately 3000
and 3 full orbits at these radii. To test the numerical convergence of
our results, we performed three additional variants of the fiducial
run, one at lower resolution, with $(\Delta r,\Delta\theta)= (40~R_s,
0.02~{\rm rad})$, and two at higher resolutions of $(\Delta
r,\Delta\theta)= (10 R_s, 0.005~{\rm rad})$, and $(\Delta
r,\Delta\theta)= (5~R_s, 0.0025~{\rm rad})$.  The corresponding
physical time--steps in these runs were $\Delta t = 70,17.5,$ and
$8.75$ seconds, respectively.

It has been shown that the black hole binary can excite eccentricities
in the disk within a region comparable to the binary separation
\citep{mm08,hayasaki07,hayasaki08,cuadra09}.  This region shrinks as
the binary separation shrinks, and our region of interest ($>10^3
R_s$) is an order of magnitude larger than the binary separation when
gravitational radiation first dominates the inspiral timescale. We
therefore assume that the gas orbits have had time to circularize, and
initially follow zero--eccentricity Kepler orbits.

The disk is assumed to consist of ideal gas, and for simplicity, we
use the constant gamma-law equation of state employed by FLASH.  We
chose the value $\gamma = 5/3$, corresponding to monatomic ideal gas.
However, the equation of state can be calculated using one of two
methods.  The first method is used in simulations we call
``adiabatic,'' which use the internal energy of each mesh zone to
calculate the temperature and pressure for the respective zone.  The
second method is ``isothermal'' for which temperature is used to
calculate the internal energy and pressure for each zone. The
temperature used in these calculations is held fixed throughout the
disk for the duration of the simulation at a pre-specified value.  In
practice, this forces $P \propto \rho$, but keeps $E_{\rm int}$
constant.  Our adiabatic and isothermal runs should be considered
limiting cases, which bracket the true behavior of a thin disk with a
more realistic treatment of radiative cooling.

For simplicity, in a first set of runs, we assume the disk initially
has a constant surface density ($\Sigma=1.5\times10^5~{\rm
  g~cm^{-2}}$; the total mass of the disk is $2.1\times10^3~{\rm
  M_\odot}$) and a constant temperature (chosen from the range
$4.5\times10^2~{\rm K} \leq T \leq 1.4\times 10^8~{\rm K}$).  These
choices are broadly motivated by the physical values expected in a
standard $\alpha$--disk around a $10^6~{\rm M_\odot}$ black hole
\citep[e.g.][]{accretionpower}.  In a second set of runs, we adopt
radial power--law profiles for the surface density, temperature, and
scale--height, given by
\begin{eqnarray}
	\Sigma(R) &=& 6.10 \times 10^{5} \left(\frac{M}{10^6~{\rm M_\odot}}\right)^{\frac{1}{5}} \left(\frac{R}{100R_s}\right)^{-\frac{27}{40}} \hspace{-0.35cm}{\rm g~cm^{-2}} \label{eq:Sigslope}\\
	T(R) &=& 5.84 \times 10^6 \left(\frac{M}{10^6~{\rm M_\odot}}\right)^{-\frac{2}{10}} \left(\frac{R}{100R_s}\right)^{-\frac{33}{40}} {\rm K} \\
	H(R) &=& 1.46 \left(\frac{M}{10^6~{\rm M_\odot}}\right)^{-\frac{1}{10}} \left(\frac{R}{100R_s}\right)^{\frac{87}{80}}R_s, \label{eq:Hslope} 
\label{eq:Tslope}
\end{eqnarray}
The scale--height $H(R)$ is not actually used in the 2D simulation,
but we use it in \S~\ref{sec:Others} below to compute the 3D gas
density, needed to predict the Bremsstrahlung luminosity. For a
standard $\alpha$--disk, with common choices for the viscosity
parameter, disk opacity, and other parameters, the power--law slopes
of the physical parameters are somewhat different in the ``inner''
(radiation pressure-- and electron scattering opacity--dominated) and
``middle'' (gas pressure-- and electron scattering opacity--dominated)
regions. The transition between these two regimes occurs around $\sim
10^3-10^4 R_s$ for typical disk parameters, but the power--law slopes
in these two regimes are, in any case, quite similar (see, e.g.,
\citealt{HKM09b}).  The power-law slopes we adopt in
equations~(\ref{eq:Sigslope})-(\ref{eq:Tslope}) are averages between
the values in these two disk regions.  Note that the total mass of our
$\alpha$--disk is $580~{\rm M_\odot}$, somewhat lower than the mass of
our uniform disk.

\begin{table}
  \caption{\it Summary of constant initial surface density disk
    simulation runs.  In all runs, the central BH has a mass of $10^6~{\rm
      M_\odot}$, and the disk extends from $10^2$ to $10^4$ Schwarzschild
    radii.  In runs that include a kick, the kick velocity is
    $530~{\rm km~s^{-1}}$, oriented in the plane of the disk, and in 
    runs that include mass loss, the fractional loss is assumed to be
    5\%.
    \label{tab:CDsims}}
\begin{center}
\begin{tabular}{l c c c}
	\hline
	\textbf{Simulation Type} & \textsl{Adiabatic} & \textsl{Isothermal}  &  Temperature \\
	\hline
	kick only & Kad1 &  Kiso1 & $450$ K \\
	(no mass loss) & Kad2 &  Kiso2 & $4500$ K \\
	 & Kad3 &  Kiso3 & $1.4\times10^5$ K \\
	 & Kad4 &  Kiso4 & $4.5\times10^5$ K \\
	 & Kad5 &  Kiso5 & $1.4\times10^6$ K \\
	 & &  Kiso6 & $1.4\times10^8$ K \\
	\hline
	Higher--resolution & & & \\ 
	\hspace{0.2 in}version of Kiso3 & -- &  Kiso3$^+$  & $1.4\times10^5$ K \\
	Highest--resolution & & & \\
	\hspace{0.2in}version of Kiso3 & -- &  Kiso3$^{++}$  & $1.4\times10^5$ K \\
	Low--resolution & & & \\
	\hspace{0.2 in}version of Kiso3 & -- & Kiso3$^-$ & $1.4 \times 10^5$ K \\
	\hline
	mass loss only & & & \\
	\hspace{0.2 in} (no kick) & Mad3 &  Miso3 & $1.4\times10^5$ K \\
	mass loss and kick & -- &  MKiso3 & $1.4\times10^5$ K \\
	\hline
\end{tabular}
\end{center}
\end{table}

\begin{table}
\caption{\it Summary of simulation runs with an initial $\alpha$--disk
profile.  As in the constant surface density runs, the central BH has
a mass of $10^6~{\rm M_\odot}$, and the disk extends from $10^2$ to
$10^4$ Schwarzschild radii; the kick velocity is $530~{\rm
km~s^{-1}}$, oriented in the plane of the disk, and the fractional
mass loss is 5\%.  The resolution in all $\alpha$--disk runs is the
same as in the highest--resolution constant--density run \textsl{Kiso3}$^{++}$.
\label{tab:AlphaSims}}
\begin{center}
\begin{tabular}{l l c}
	\hline
	 & \textbf{Simulation Type} &  \textbf{Abbreviation} \\
	\hline
	\textsl{Isothermal} & &  \\
	 & kick only (no mass loss) & Kiso \\
	 & mass loss only (no kick) & Miso \\
	 & kick and mass loss & MKiso \\
	\textsl{Adiabatic} & & \\
	 & mass loss only (no kick)  & Mad \\
	 & kick and mass loss & MKad \\
	\hline
\end{tabular}
\end{center}
\end{table}

The presence of the circumbinary disk is expected to align the spins
of both BHs during the early stages of their evolution (at large
separations), to be parallel with the angular momentum of the disk
\citep{Bogdanovic}.  In this case, the recoil will be oriented within
the plane of the disk; in any case, the two--dimensional nature of our
simulations precludes us from studying out-of-the plane kicks.  We
further use the typical value of 530 ${\rm km~s^{-1}}$ \citep{baker08}
throughout this paper.  Our simulations are set up in the reference
frame traveling with the kicked SMBHB.  Since the total disk mass
within our simulated region of $100-10,000 R_s$ is only a fraction
$\sim 10^{-3}$ of the SMBHB mass, we neglect the inertia of the gas
bound to the SMBHB. In practice, we simply add a unidirectional
velocity component to the gas, everywhere in the plane of the disk, in
our initial conditions, on top of the circular Kepler velocities.  In
runs which include a mass loss, we adopt a fractional loss of 5\% --
i.e, the disk is set up as before, but the central mass is reduced to
$9.5\times 10^5~{\rm M_\odot}$ at the first time--step of the
simulation.

Tables \ref{tab:CDsims} and \ref{tab:AlphaSims} summarize the
parameters of all of our simulations.  The runs were performed at the
Center for Cosmology and Particle Physics at New York University using
the Ria cluster.  Ria has 368 2.6 GHz CPUs with Infinband interconnect
and 768~GB of memory.  For reference, we note that our fiducial run,
labeled \textsl{Kiso3}, was run on three nodes (24 processors), and
took four days to complete.

\section{Results and Discussion}
\label{sec:Results}

\subsection{Post-Merger Disk Evolution and Shock Structure}
\label{sec:PostMerger}

\begin{figure*}
\centering
	\includegraphics [scale=0.6] {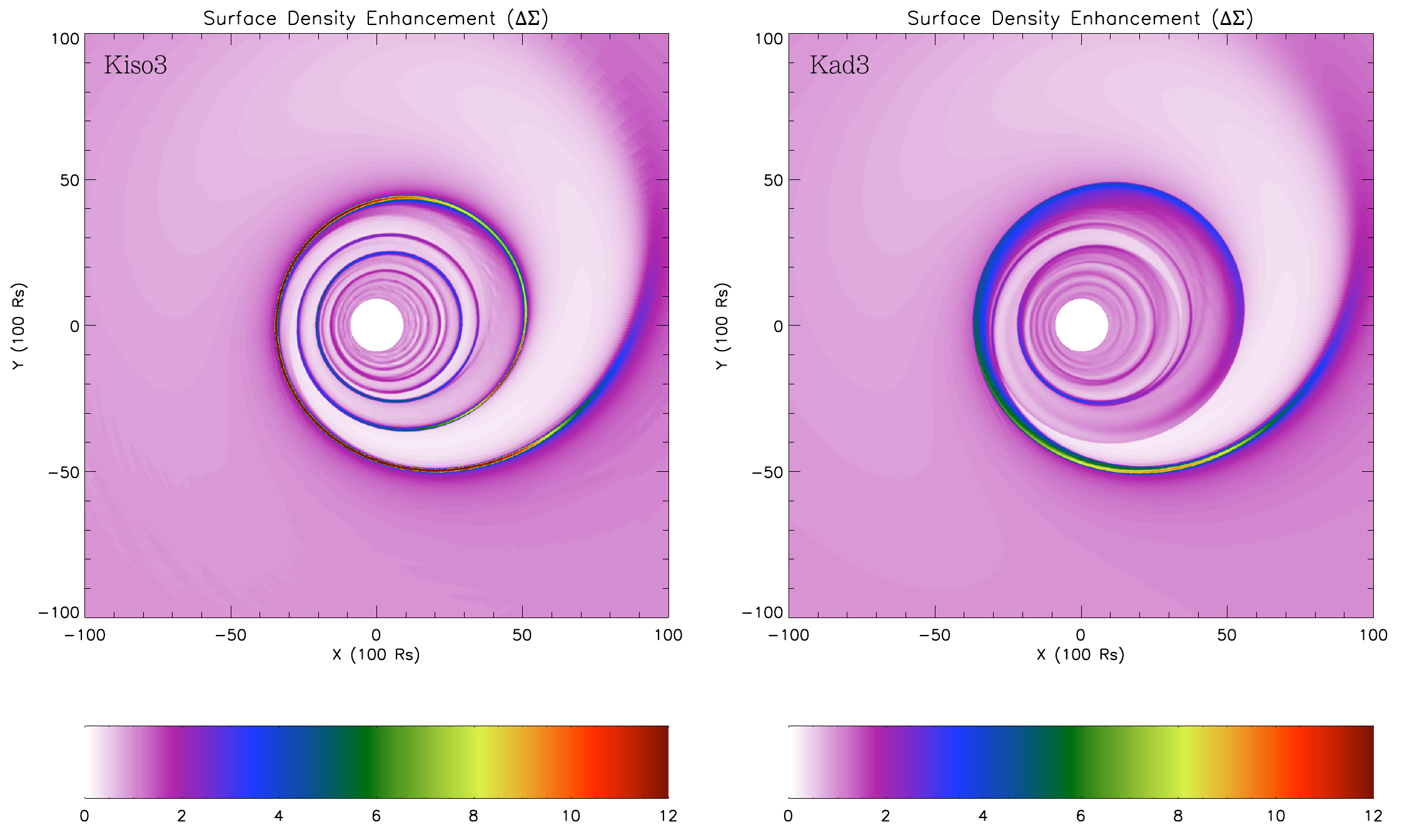}
	\caption{Contour images of the surface density distributions
          in our fiducial isothermal (\textsl{Kiso3}; left) and
          adiabatic (\textsl{Kad3}; right) simulations at $t=210$ days
          following the merger and the recoil of the SMBHB.  In both
          cases, the rotation of the disk is clockwise, and the
          direction of the kick is up along the $y$ axis; the disks
          are shown in Cartesian coordinates extending to $10^4
          R_s$. The color scales at the bottom are in units of the
          overdensity, relative to the constant initial surface
          density.}
	\label{fig:diskdens}
\end{figure*}

Figure~\ref{fig:diskdens} shows snapshots of the two--dimensional
surface density distribution in our fiducial isothermal
(\textsl{Kiso3}; left panel) and adiabatic (\textsl{Kad3}; right
panel) simulations $t=210$ days after the merger.  As the figure
shows, sharp overdensities develop in both cases, with a morphology
similar to the spiral caustics in the collisionless case
\citep{Lippai}.  The density contrasts are visibly higher in the
isothermal case, as one intuitively expects from the higher
compressibility of isothermal gas.

The sharpness of the thin arcs traced out by the over-dense regions
suggest the presence of shocks.  To assess whether shocks have indeed
formed, once the simulation runs have completed, the output files were
tested with an algorithm identical to the {\it shock\_detect} feature
available in FLASH.  To receive a shock flag, a zone must pass both of
the following criteria.  First, the pressure difference between the
zone and at least one of its neighbors must exceed the difference
expected from the Rankine-Hugoniot jump conditions for a shock of a
pre--specified minimum Mach number $\mathcal{M}$.  To allow for
oblique shocks, while comparing each zone to its neighbors in
different directions, the pressure is weighted according the fraction
of the total velocity made up by the velocity vector in the
appropriate direction.  Second, to avoid falsely flagging expanding
regions as shocks, the velocity divergence at the zone in question
must be negative.

The zones identified by the above criteria shows that shocks indeed
arise in our runs at early times, and propagate outward.  Shocks at
early times in the disk were found to be typically weak.  For example,
in our \textsl{Kiso3} run, at $t=30$ days, among the shocked zones
identified with the threshold $\mathcal{M} > 1.1$, we find that 20\%
exceeded $\mathcal{M} > 2$. Subsequently, the shocks increased in
strength, as they moved outward (with the top 20\% of the shocked
zones having $\mathcal{M} > 3$ by $t=200$ days).  Shocks appeared in
all of our runs except \textsl{Kiso6} -- in this run, the constant
isothermal temperature was increased to correspond to a sound speed of
$c_s \approx 10^3~{\rm km~s^{-1}}$, approximately two times larger
than $v_{\rm kick}$.  This suggests that $v_{\rm kick}\gsim c_s$ can
be used as an approximate criterion for the kick to cause shocks.
This is consistent with the expectation in collisionless disk, in
which particles cross their epicyclic orbits with a relative velocity
of $\Delta v \sim v_{\rm kick}^2/v_{\rm orbit}$; in bound regions of
the disk, this relative velocity is limited to $\Delta v \lsim v_{\rm
  kick}$ \citep{Lippai}.

\begin{figure}
\centering
	\includegraphics [scale=0.55] {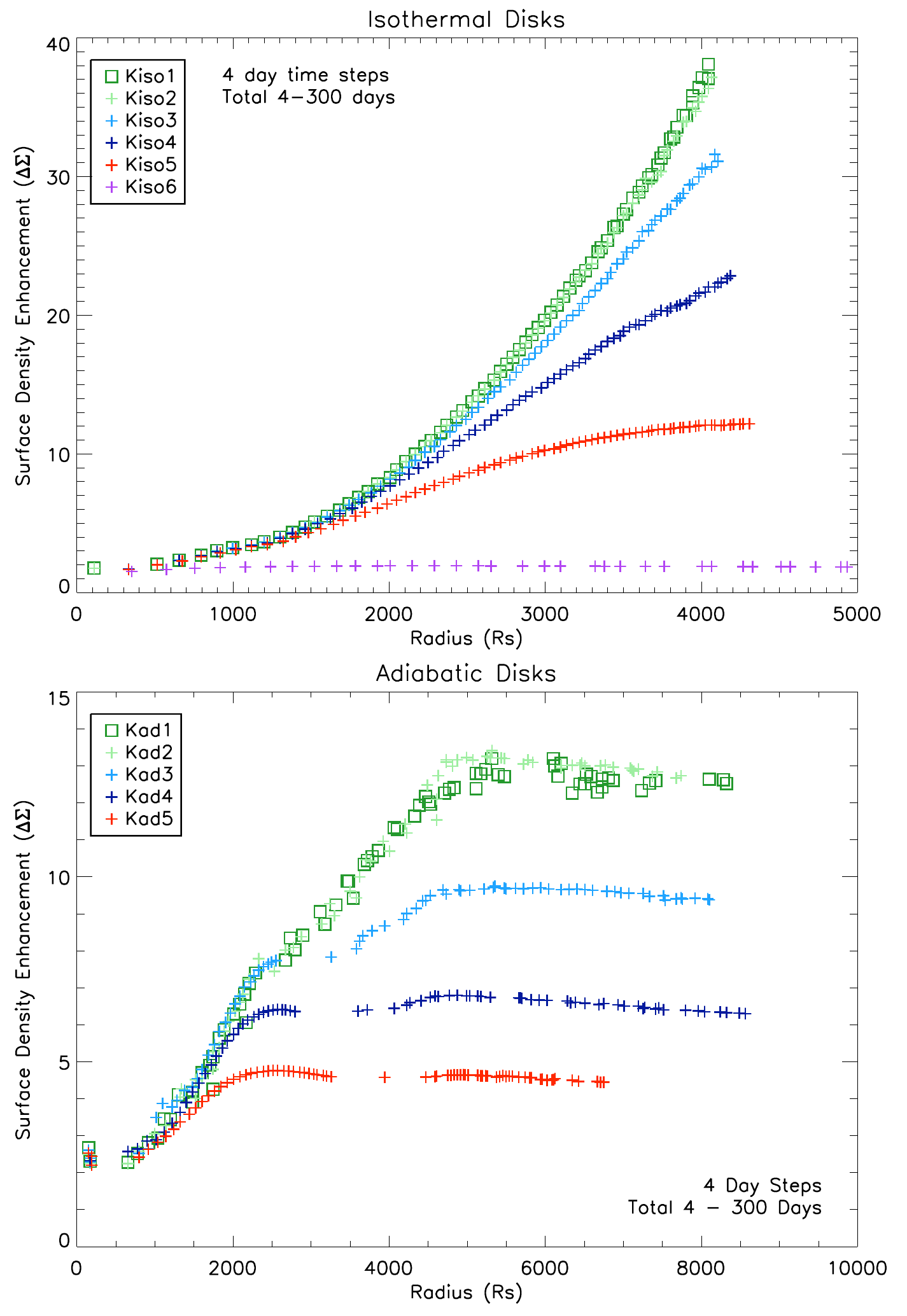}
	\caption{Evolution of the maximum surface density in our
          isothermal (\textsl{Kiso1-6}; upper panel) and adiabatic
          (\textsl{Kad1-5}; lower panel) runs.  The points show the
          radial position and the local overdensity at the densest
          point in the disk, identified at several different times
          between $t=4$ and $t=300$ days after the merger (with 4 days
          corresponding to the left--most point, and 300 days to the
          right--most point).  }
	\label{fig:DensRad}
\end{figure}

In Figure~\ref{fig:DensRad}, we show the radial position of the
densest point identified at several different times in the isothermal
(\textsl{Kiso1-6}; upper panel) and adiabatic (\textsl{Kad1-5}; lower
panel) runs, together with the surface density enhancement at this
point, defined as $\Delta\Sigma\equiv \Sigma/\Sigma_i$ (where
$\Sigma_i$ is the initial constant surface density).  The points are
shown every 4 days, between $t=4$ and $t=300$ days after the merger,
with 4 days corresponding to the left--most point, and 300 days to the
right--most point.  In the isothermal runs the densest point in the
shock wave tends to stay at the same azimuthal position in the disk,
$\sim 3 \pi / 2$ rad (as measured clockwise from the y axis), which
corresponds to the region of the disk where the kick direction is
parallel to the tangential velocity of gas in the disk.  The kick adds
constructively to the Keplerian velocity of gas at the angle $3
\pi/2$, and it is not surprising that gas at this azimuthal position
achieves the largest over-densities first.

\begin{figure}
\centering
	\includegraphics [scale=0.45] {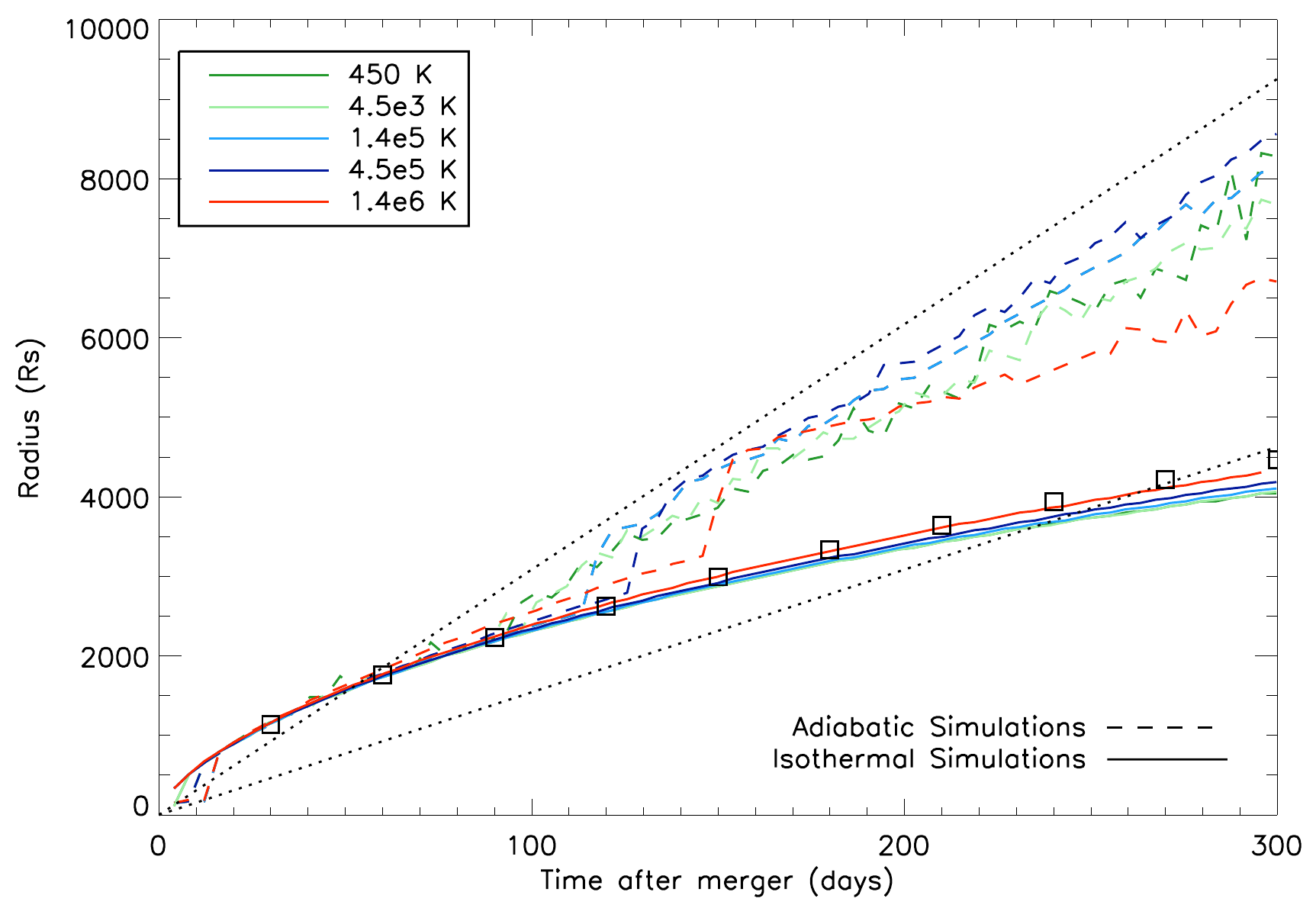} \\
        \caption{The figure shows the radial location of the densest
          point in the disk as a function of time. The same runs are
          shown as in Figure~\ref{fig:DensRad}, with the same notation
          (except \textsl{Kiso6} is excluded, since it had no shocks).
          For comparison, the black squares show the location of the
          outermost caustic identified in a collisionless disk
          \citep{Lippai}.  The straight dotted lines correspond to
          propagation at constant speeds of $v_{\rm kick}=530~{\rm
            km~s^{-1}}$ (lower line) and $2v_{\rm kick}=1060~{\rm
            km~s^{-1}}$ (upper line).}
	\label{fig:Rshock}
\end{figure}

In Figure~\ref{fig:Rshock}, we show the radial position of the densest
point in the disk as a function of time, in the same 10 runs as in
Figure~\ref{fig:DensRad} (\textsl{Kiso6} is excluded).  For
comparison, the black squares show the location of the outermost
caustic identified in a collisionless case (\citealt{Lippai}; see
further discussion in \S~\ref{sec:Others} below).  The location of the
isothermal shocks follow the caustics quite closely.  However,
interestingly, after $t\approx 100$ days, the densest shock appears to
propagate $\sim$ twice as rapidly in the adiabatic simulations. As
Figure~\ref{fig:diskdens} shows, the overall size and morphology of
the sharp spiral patterns forming in the isothermal and adiabatic runs
are quite similar, except that in the adiabatic runs, the spiral arms
are more diffuse and each arm tends to extend to a somewhat larger
radius than in the corresponding isothermal case.  The difference in
the apparent radial propagation speed of the densest point arises
because typically, in the adiabatic runs (i) the densest point falls
further out along the outermost spiral arm, and (ii) the outermost
spiral arm extends to slightly a larger radius.  Furthermore, the
densest point is typically farther out along the spiral pattern in
adiabatic disks with increasing temperatures.  Interestingly, the
adiabatic run with the hottest temperature (\textsl{Kad5}) is an
exception; in this run, the densest point again falls further in along
the outermost spiral arm, which is noticeably more diffuse compared to
the \textsl{Kad4} run.  The spiral arms in the adiabatic runs also
tended to diffuse over time, with pairs of adjacent arms eventually
overlapping. In comparison, the isothermal spiral arms remained
distinct and did not interact.  This likely explains why the
isothermal curves are much smoother than the adiabatic curves.

These figures also illustrate the strength of the density
perturbations caused by the kick, allowing us to draw several
interesting conclusions.  In particular, as the temperature, and
therefore the pressure increases, the density enhancements generally
decrease, as expected.  The two lowest temperature cases ($T=450$ and
$4500$K) differ very little in either the adiabatic or isothermal
case, indicating that pressure effects are negligible when the disk
temperature is below $T \approx 4500 K$.  Also as expected, the
overdensities are much larger in the isothermal runs, reaching
$\Delta\Sigma\gsim 40$ in our fiducial \textsl{Kiso3} case, and still
rising at the end of the run, whereas in the adiabatic run with the
same temperature (\textsl{Kad3}), the overdensity reaches a plateau
and is limited to $\Delta\Sigma < 10$.  Pressure has little effect at
very early times (except for the extremely hot disk \textsl{Kiso6}),
when the shocks are within $\lsim 10^3 R_s$ and have relatively
low--overdensities. Note that the sound crossing time near the inner
edge of the disk in our fiducial run is $R/c_s\approx 70$ days,
comparable to the time when pressure starts to have an impact on the
shock overdensities.  Interestingly, pressure suppresses density
enhancements at late times by large factors.  This last conclusion is
somewhat surprising, given expectations from collisionless disks (see
further discussion of this point in \S~\ref{sec:Others} below).

\begin{figure}
\centering
	\includegraphics [scale=0.45] {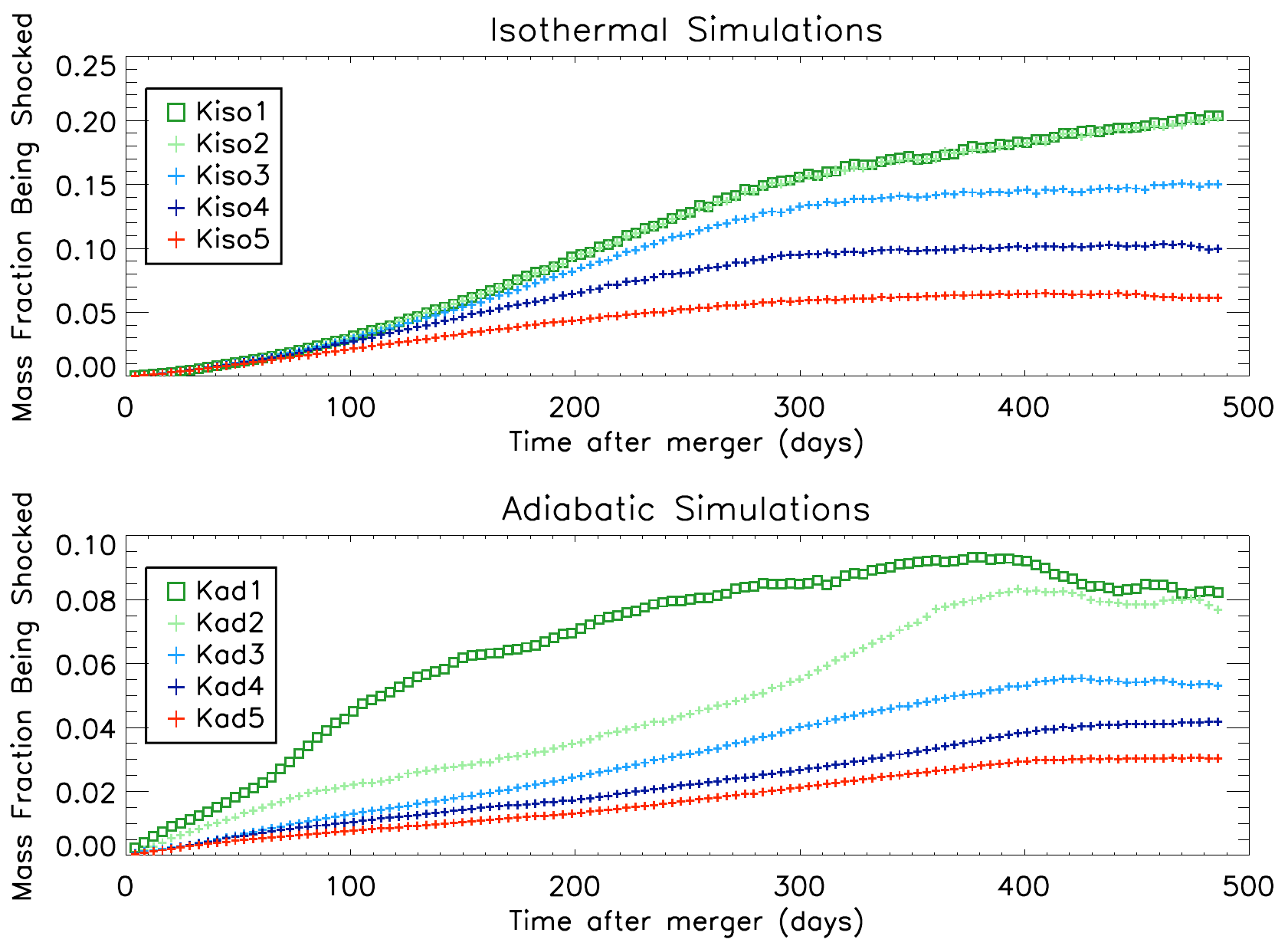}
	\caption{Mass fraction of the disk material undergoing a shock
          with a minimum Mach number of $\mathcal{M} \geq 1.1$, as a
          function of time. The different symbols correspond to the
          runs shown in Figure~\ref{fig:DensRad}.  The curves begin to
          flatten when the outermost part of the spiral pattern begins
          to leave the high--resolution mesh region, in both the
          isothermal ($\sim 250$ days) and the adiabatic simulations
          ($\sim 400$ days; note that in the adiabatic runs, a larger
          mesh was used, extending to $15,000R_s$).}
	\label{fig:PercentShock}
\end{figure}

In Figure~\ref{fig:PercentShock}, we show the total mass of the
shocked material divided by the total mass of the disk at a given
time, all within the high--resolution mesh region ($100 < R/R_s <
10^4$ for \textsl{Kiso1-5}, and $100 < R/R_s < 1.5\times10^4$ for
\textsl{Kad1-5}) using the shock criteria defined above, and a minimum
Mach number of $\mathcal{M}=1.1$.  For cold isothermal disks, a
significant fraction of the disk ($\sim 20\%$) can be experiencing a
shock at a given moment.  The shocks in the adiabatic cases comprise a
smaller fraction of the disk, because the temperature and pressure
($\propto \rho T$) both increase in the compressed regions, reducing
the bulk velocities and shock strengths.  Overall, compared to the
isothermal cases, in the adiabatic runs the density enhancements are
reduced, and shocks are weakened and are less prevalent -- these
trends are as expected, and follow from the lower compressibility of
adiabatic gas.

\subsection{Impact of Mass-Loss}
\label{sec:MassLoss}

The results in the previous section included only the effect of the
kicks, and not the effects of the accompanying mass loss.  Upon
coalescence of the SMBH binary, energy is carried away by
gravitational waves, causing an instantaneous reduction of the rest
mass by the amount of $(M_1 + M_2) \epsilon_{GW}$.  The fractional
loss depends on the mass ratio, on the orbital parameters, and on the
magnitude and orientation of both BH spins, for which the full
parameter space has not yet been explored in direct relativistic
numerical simulations.  For nearly equal mass binaries ($q\equiv
M_1/M_2 \sim 1$), however, the mass loss can be several percent
\citep[e.g.][]{tm08,herr07a}.  For simplicity, throughout this paper,
we adopt a constant fractional mass loss of 5\%.  We initialize the
disk with Keplerian velocities for the original pre--merger point mass
$(M_1+M_2)=10^6~{\rm M_\odot}$.  However, in the remainder of the
simulation, the gravitational potential is that of a central point
mass with $(M_1+M_2) (1-\epsilon_{GW})=9.5\times10^5~{\rm M_\odot}$.

\begin{figure*}
\centering
	\includegraphics [scale=0.6] {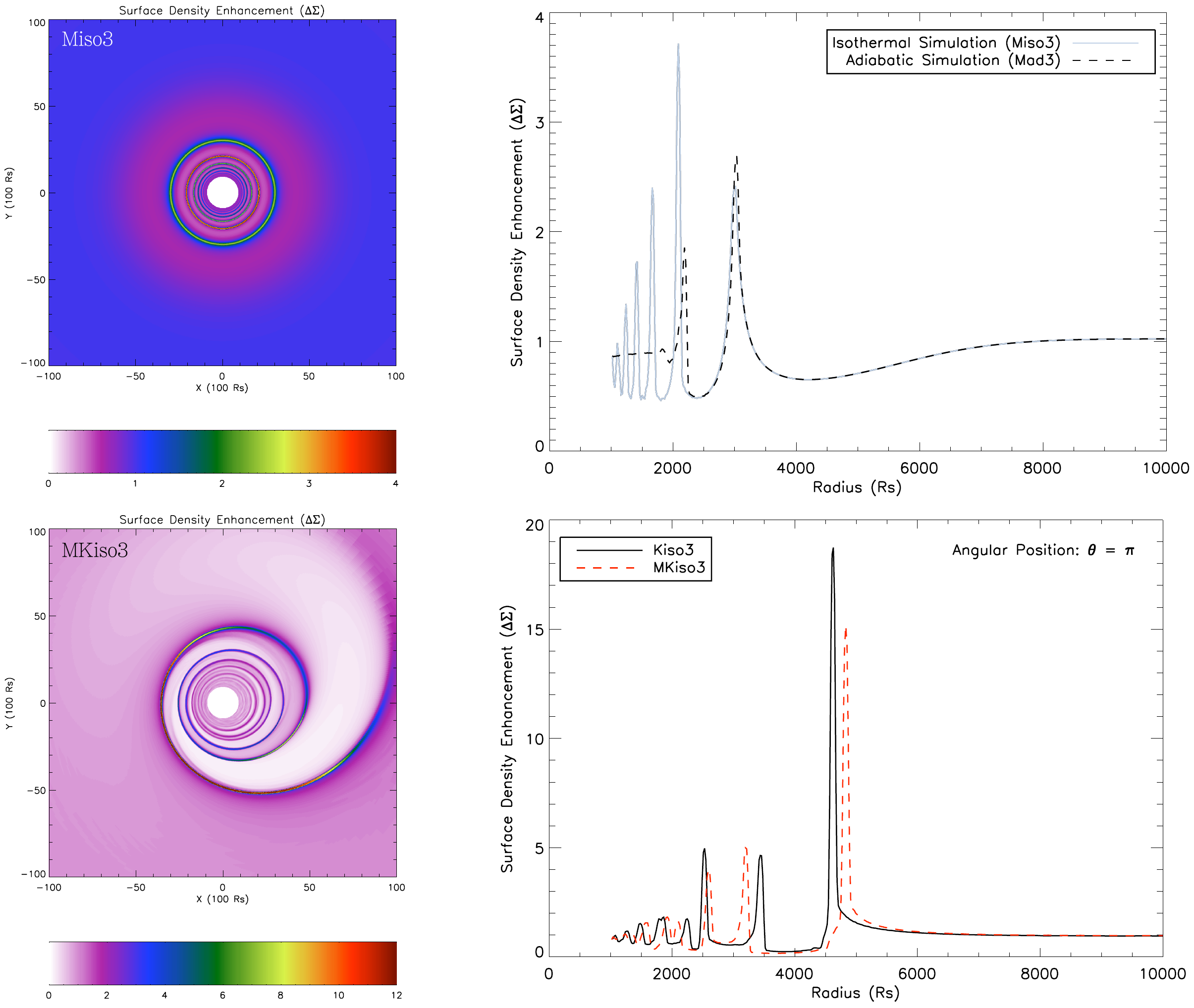}
	\caption{{\it Top panel:} The left-hand side contour map shows
          the surface density for an isothermal $1.4 \times 10^5$ K
          disk, where the central BH experiences sudden mass--loss by
          5\% (\textsl{Miso3}). No kick is included. Concentric rings
          of density spikes develop and propagate outward, here shown
          at $t=210$ days after the merger.  The figure on the right
          shows the radial profile of the disk surface density for the
          isothermal disk (\textsl{Miso3}) in comparison to surface
          density from the adiabatic disk (\textsl{Mad3}) which was
          subjected to the same initial setup.  {\it Bottom panel:}
          The contour map on the left shows the surface density of a
          disk at $t=210$ days after the merger.  A kick, with $v_{\rm
            kick}=530~{\rm km~s^{-1}}$ is included in addition to the
          mass loss (\textsl{MKiso3}).  In the right panel, we also
          show, for comparison, the radial overdensity profile in our
          isothermal disk that includes only a kick, and no mass loss
          (\textsl{Kiso3}).  Interestingly, the addition of the
          mass--loss results in a small reduction of the kick--induced
          overdensities (as well as a slightly faster
          outward--propagation of the outermost shock). }
	\label{fig:MassLoss}
\end{figure*}

The effects of such an instant mass-loss was treated recently by
\citet{ONeill} and \citet{Megevand} in three--dimensional adiabatic
disks.  Here we extend these works by following the impact of the
mass-loss case in an adiabatic (\textsl{Mad3}) as well as in an
isothermal disk (\textsl{Miso3}).  In the latter case, we also
investigate the combined impact of mass loss and recoil
(\textsl{MKiso3}).  As noted in the Introduction, having one less
dimension allows our simulations to cover the disk out to a much
larger radius ($10^2-10^4 R_s$ vs. the range $2-50 R_s$ in
\citealt{ONeill}), and for a much longer duration ($\sim 1$yr
vs. $\sim 5$ hrs in \citealt{ONeill}; for a $10^6~{\rm M_\odot}$ BH).

The top portion of Figure~\ref{fig:MassLoss} shows a surface density
map of an isothermal disk, in the mass--loss case, around $t=210$ days
after the merger (in the left panel).  In the right panel of the same
figure, we show the radial surface density profile of the same disk,
together with the profile of an adiabatic disk with the same
mass--loss.  When the impact of the kick is not included, the
perturbations are azimuthally symmetric, giving rise to concentric,
outward--propagating ripples.  The basic mechanism behind creating the
ripples can be visualized simply as follows: at the moment of mass
loss, each particle, orbiting at its original Keplerian velocity, is
moving at a speed that is too fast for a circular orbit around the
reduced central mass.  As a result, all particles find themselves at
the pericenter of a new, elliptical orbit. Summing the effective
contributions of each particle to the surface density, spread over
these orbits, results in a net {\it dilution} of the disk.
Furthermore, the radial epicyclic oscillations of the set of particles
at a fixed initial radius will occur at a frequency that monotonically
decreases radially outward, resulting in a pattern of quasi--periodic
compression / decompression in the radial direction.  The resulting
density spikes have lower amplitudes, but are more frequent toward the
inner boundary.  These effects are clearly seen in
Figure~\ref{fig:MassLoss} except that, notably, the pressure forces in
the adiabatic case are sufficient to almost completely wash out the
innermost density ripples, out to $R\lsim 2,000 R_s$, by $t=210$ days.

When a kick is included in addition to the mass-loss
(\textsl{MKiso3}), the density enhancements are often lower in
comparison to the kick-only case (\textsl{Kiso3}).  The bottom portion
of Figure~\ref{fig:MassLoss} shows a map of the surface density and a
radial slice of the surface density profile along the azimuth
$\theta=\pi$ (running in the direction of the -$y$ axis) in both
cases.  The density enhancement from the mass--loss + kick run is
appreciably lower than in the kick-only case, especially for the
outermost, densest spike, at $R\approx 4,500-5,000 R_s$.

\subsection{Observational Signatures}
\label{sec:Observational}

Our adiabatic runs conserve energy, so, by definition, no energy loss
occurs in these runs, and the disk has strictly zero luminosity.  In
reality, the disk will radiate -- indeed, a prediction of its
time--variable light--curve, following the merger, will be invaluable
for identifying EM counterparts to {\it LISA} sources.  A realistic
prediction requires solving for the vertical structure of the disk, as
well as a treatment of radiative cooling and radiative transfer across
the disk (which is likely to be optically thick; see below).  While
this is not possible in our 2D simulations, our isothermal runs allow
us to calculate the energy dissipation occurring in the disk -- this
can serve as a rough, order--of--magnitude estimate for the disk
luminosity.

More specifically, at each time--step, FLASH solves Euler's equations,
expressed in conservation form, to update the total specific energy
($E$) of each zone which includes the combined kinetic and internal
energy.  In our case, the disk is cold, and the internal energy
($E_{\rm int}$) is very small compared to the kinetic energy of
orbital motion, and is therefore a small fraction of the total energy
$E$. This can cause numerical inaccuracies, and we found that it was
necessary to solve for $E_{\rm int}$ explicitly, using an internal
energy advection equation that is separate from the total
energy\footnote{See equation 13.7 in the FLASH User Guide, Version
  3.2, July 2009; available at
  http://flash.uchicago.edu/website/codesupport/flash3$\_$ug$\_$3p2.pdf.}.
The pressure (and temperature, if required) of the zone is
subsequently obtained using the specified equation of state.

When an isothermal equation of state is used, the pre--specified
constant initial temperature $T^{\rm iso}$ is used to calculate the
internal energy, so that it is reset to $E_{\rm int}^{\rm iso} =
(\gamma-1)^{-1} \rho k T^{\rm iso}/\mu m_p\propto \rho$.  The
difference of the internal energies before and after this re-set,
$\Delta E_{\rm int} = E_{\rm int} - E_{\rm int}^{\rm iso}$,
corresponding to the $p\Delta V$ work done by the gas, is effectively
radiated instantaneously; we compute the corresponding luminosity by
dividing $\Delta E_{\rm int}$ by the simulation time-step.

The light curve obtained by this method corresponds to the limiting
case of efficient radiative cooling that keeps the gas at constant
temperature.  An ambiguity in this interpretation arises, however, in
zones which expand, and therefore would cool, in the adiabatic case.
The isothermal condition requires that these fluid elements gain,
rather than loose energy, in order to maintain their constant
temperature.  The above calculation simply returns a negative
luminosity in these zones, which should either be excluded from the
total luminosity budget (if, in reality, these zones remain cooler
than $T^{\rm iso}$), or may be included (under the assumption that the
corresponding heating comes at the expense of the radiation generated
in the compressed regions). Clearly, the isothermal approach is not
adequate by itself to resolve this ambiguity, and below we will simply
quote luminosities both with and without the negative contribution of
these expanding zones.  These may also be regarded as upper and lower
limits; our light--curves are inherently limited to an accuracy
comparable to the difference between these two estimates (which we
find is typically a factor of $\sim$ two).

\begin{figure}
\centering
	\includegraphics [scale=0.45] {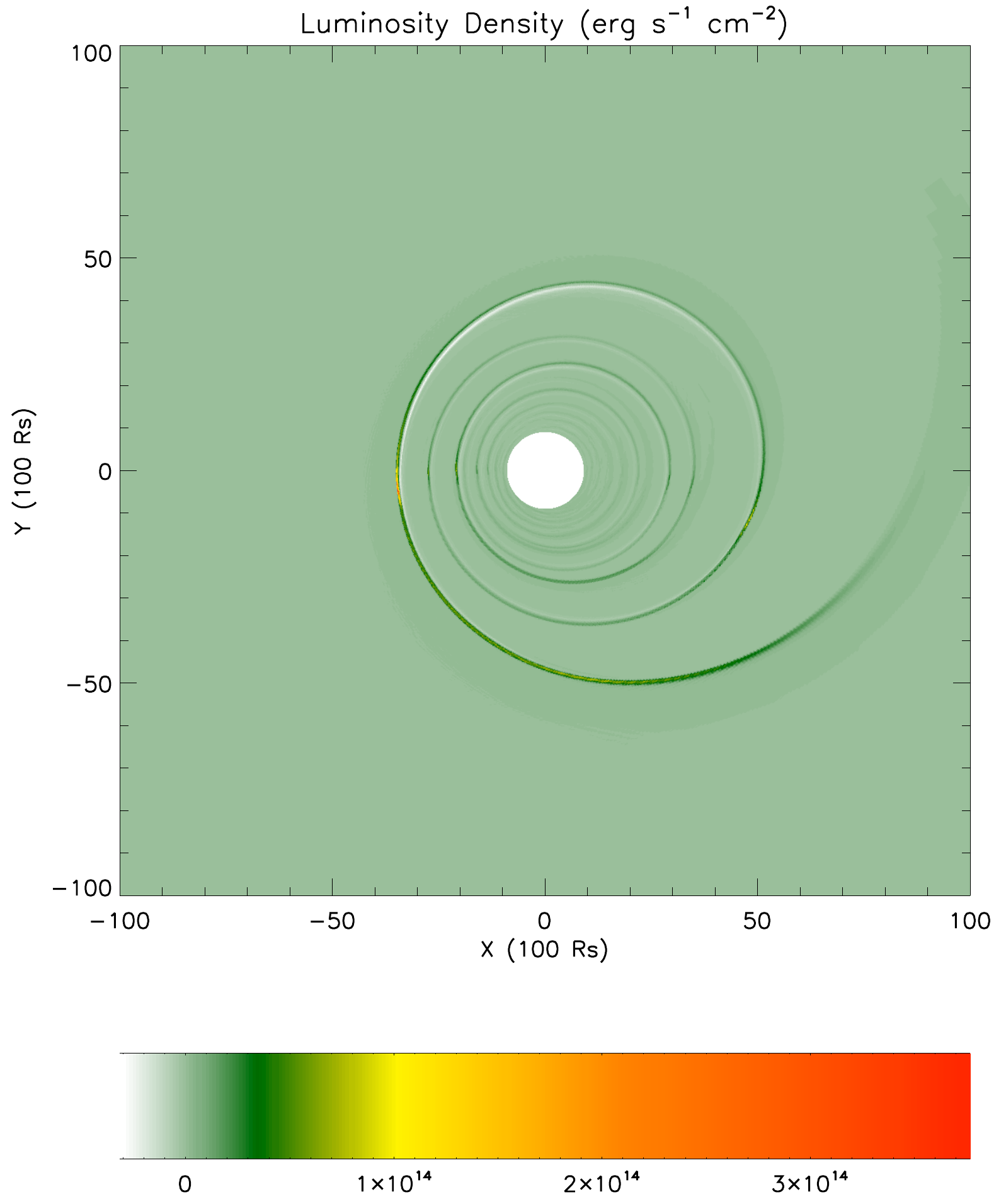}
	\caption{A snapshot of the luminosity density, defined as the
          the energy radiated per unit area and unit time, in our
          fiducial \textsl{Kiso3} run, here shown at $t=210$ days.
          The bright regions, with a high positive luminosity, closely
          trace the spiral shocks seen in
          Figure~\ref{fig:diskdens}. The color scale shown at the
          bottom of the figure is in units of ${\rm
            erg~s^{-1}~cm^{-2}}$. Note the negative contributions from
          the post-shock regions (see text for discussion).}
	\label{fig:Dei2D}
\end{figure}

Figure~\ref{fig:Dei2D} shows a snapshot of the luminosity density,
defined as the energy radiated per unit area in a single time--step
$\Delta t$, $j\equiv\Delta E_{\rm int}/\Delta A \Delta t$ in our
fiducial \textsl{Kiso3} run, at $t=210$ days.  As noted above, the
luminosity in the region inside $R<10^3R_s$ is dominated by numerical
noise. To be conservative, we have therefore excised these inner
regions from our luminosity calculations.  Outside $R<10^3R_s$, the
bright regions, with a high positive luminosity, clearly trace the
shocks quite well, which is indeed what one expects, since these are
the locations where the gas is being most compressed.  Adjacent to the
bright shocks, the figure also shows regions of post--shock
decompression, where the velocity divergence is positive, and the
luminosity is negative.

\begin{figure}
\centering
	\includegraphics [scale=0.45] {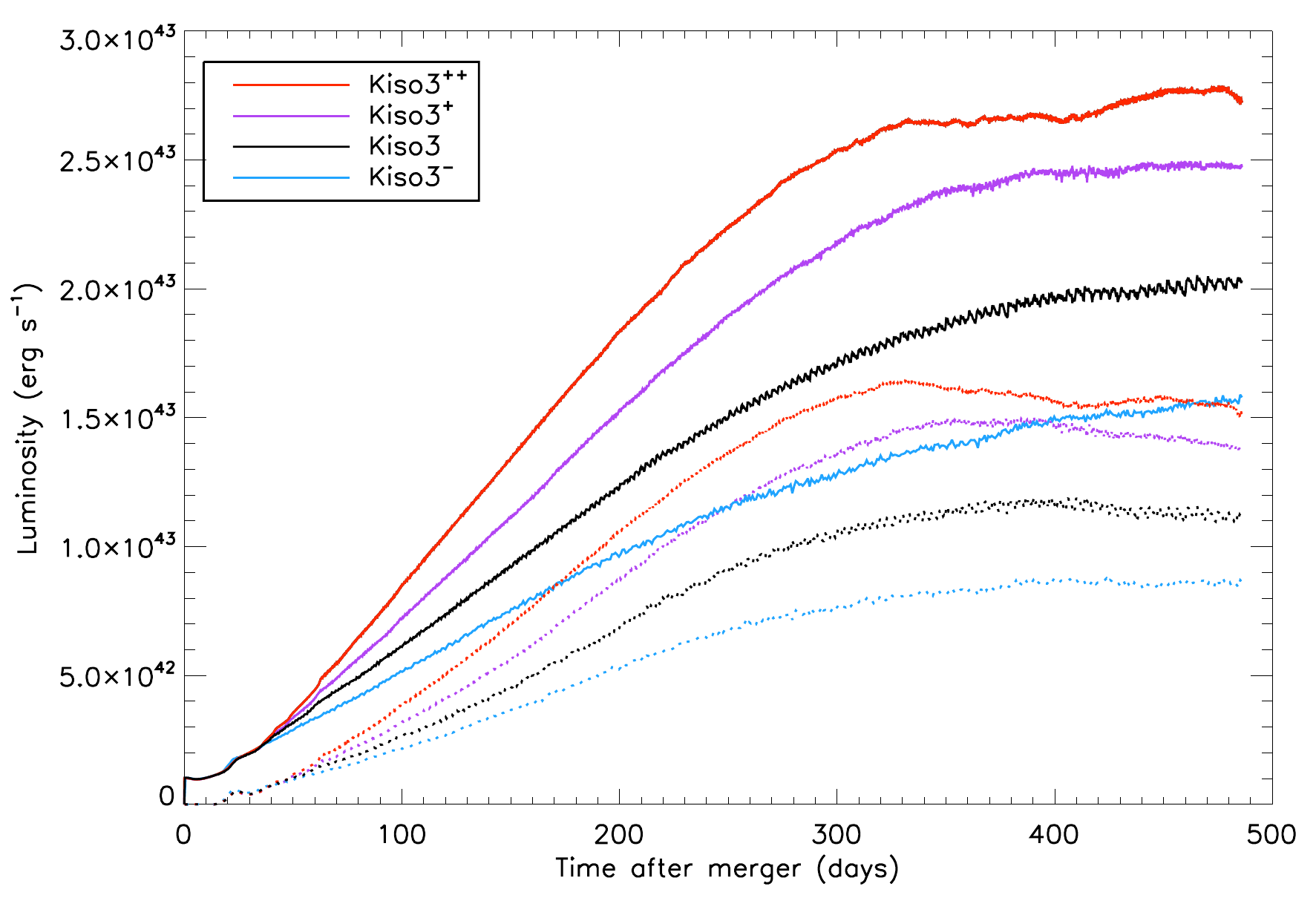}
	\caption{Light--curves obtained in $T=1.4\times10^5$K
          isothermal disks experiencing a kick, by integrating the
          luminosity density, such as shown in Fig.~\ref{fig:Dei2D},
          across the surface of the disk.  The solid curves include
          only the positive--luminosity zones, whereas the dotted
          curves include the contribution from the
          negative--luminosity zones.  The black curves are derived
          from our fiducial run (\textsl{Kiso3}), and the
          magenta/red/blue curves show the results of a resolution
          study (corresponding to the high--resolution runs
          \textsl{Kiso3}$^{++}$ and \textsl{Kiso3}$^{+}$; and the
          low--resolution run \textsl{Kiso3}$^{-}$, respectively).
          The total positive luminosity in the highest--resolution run
          (top curve) reaches $\approx$20\% of the Eddington limit for
          a $10^6 M_\odot$ black hole.}
	\label{fig:LightCurves}
\end{figure}

In Figure~\ref{fig:LightCurves}, we show the corresponding light
curves in our fiducial run (\textsl{Kiso3}) -- again, conservatively,
limited to the region $10^3 < R/R_s < 10^4$ -- when only the
positive--luminosity zones are considered (black solid curve) and when
the contribution from the negative--luminosity zones is subtracted
(black dotted curve).  The other curves in this figure show how the
light--curves depend on numerical resolution, and will be discussed in
\S~\ref{sec:Numerical} below. The absolute value of the luminosity is
significant, corresponding to $\approx 16\%$ of the Eddington
luminosity $L_{\rm Edd}$ for a $10^6 M_\odot$ black hole in the first
case, and reduced by less than a factor of two, to $\approx 10\%$
$L_{\rm Edd}$ in the latter case.  We note that despite the
significant luminosity, the total energy radiated in the first year
after the merger (integrating the light curve in
Figure~\ref{fig:LightCurves}) is found to be $\approx 0.1\times M_{\rm
  disk} v_{\rm kick}^2$.  In principle, the kick--induced dissipation
could extract and radiate away a non--negligible fraction of the total
potential energy of the disk; the luminosity we find implies an energy
release well below this upper limit.

\begin{figure}
\centering
	\includegraphics [scale=0.45] {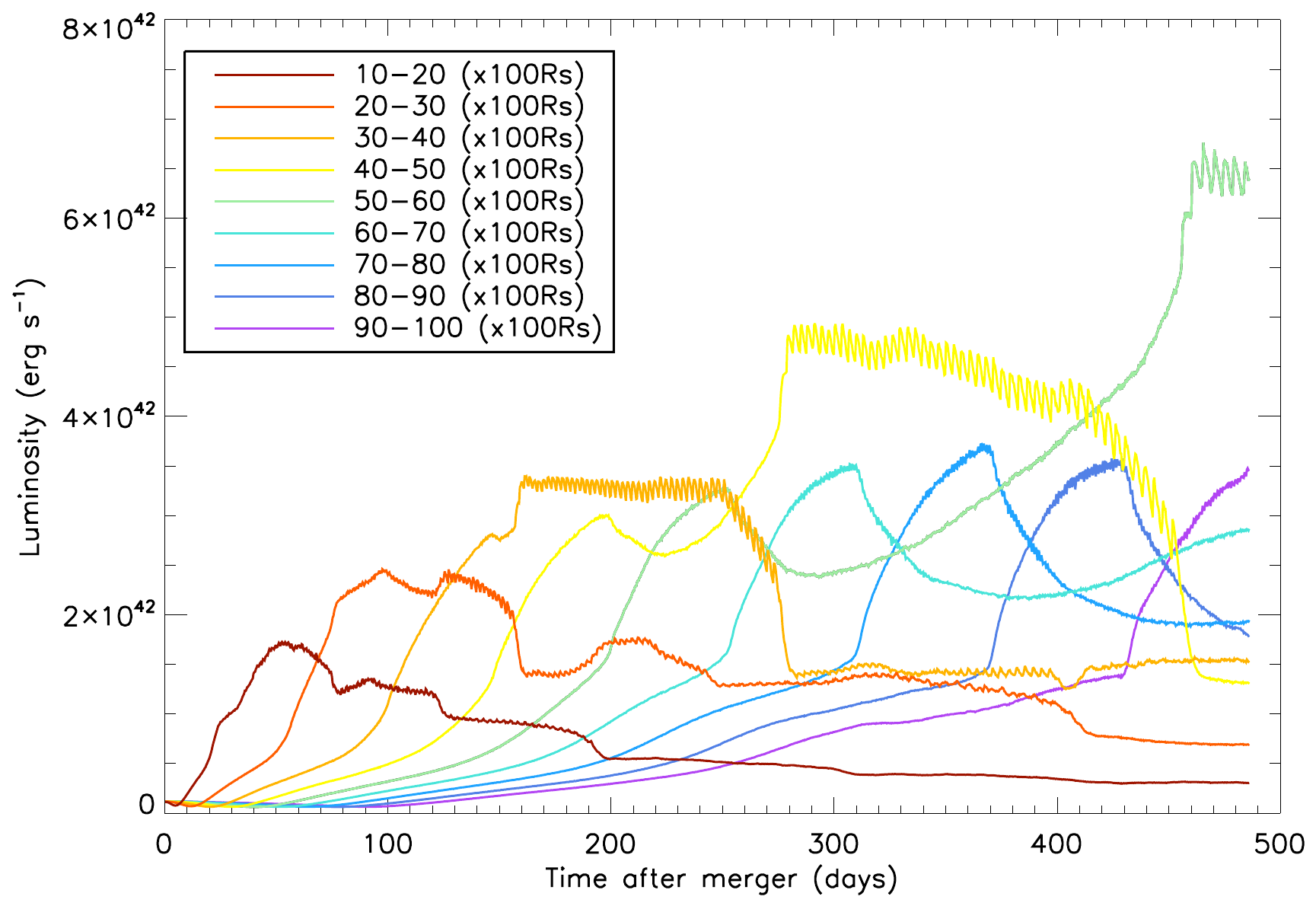}
	\caption{Contributions to the total (positive) luminosity from
          different annuli in the fiducial run \textsl{Kiso3}, as
          labeled.  The densest shocked region of the disk tends to
          dominate the total luminosity, and each annulus provides a
          maximum contribution when this ``bright spot'' moves across
          it.}
	\label{fig:BinLumi}
\end{figure}

In Figure~\ref{fig:BinLumi}, we decompose our light--curve, and show
the contribution of the positive--luminosity zones from different
annuli in the disk, binned by radius.  The densest shocked region of
the disk tends to dominate the total luminosity, and, as the figure
shows, each annulus provides a maximum contribution when the ``bright
spot'' moves across it.  The luminosity rises steadily overall, which
is attributable predominantly to having more radiating material at
larger radii (the bins shown in Fig.~\ref{fig:BinLumi} have equal
radial widths, so their area grows linearly with radius).  As
discussed above, the shocks also get stronger with time; however, this
is a relatively weak effect, and the overall rise in the luminosity,
seen in Figure~\ref{fig:LightCurves}, roughly tracks the build-up of
the shocked mass (see Fig.~\ref{fig:PercentShock}).  At the most
luminous portion of the shock, in our fiducial run, the temperature
typically rises during each time step by $\approx 4\%$, before it is
re--set to $1.4 \times 10^5$ K.

\begin{figure}
\centering
	\includegraphics [scale=0.45] {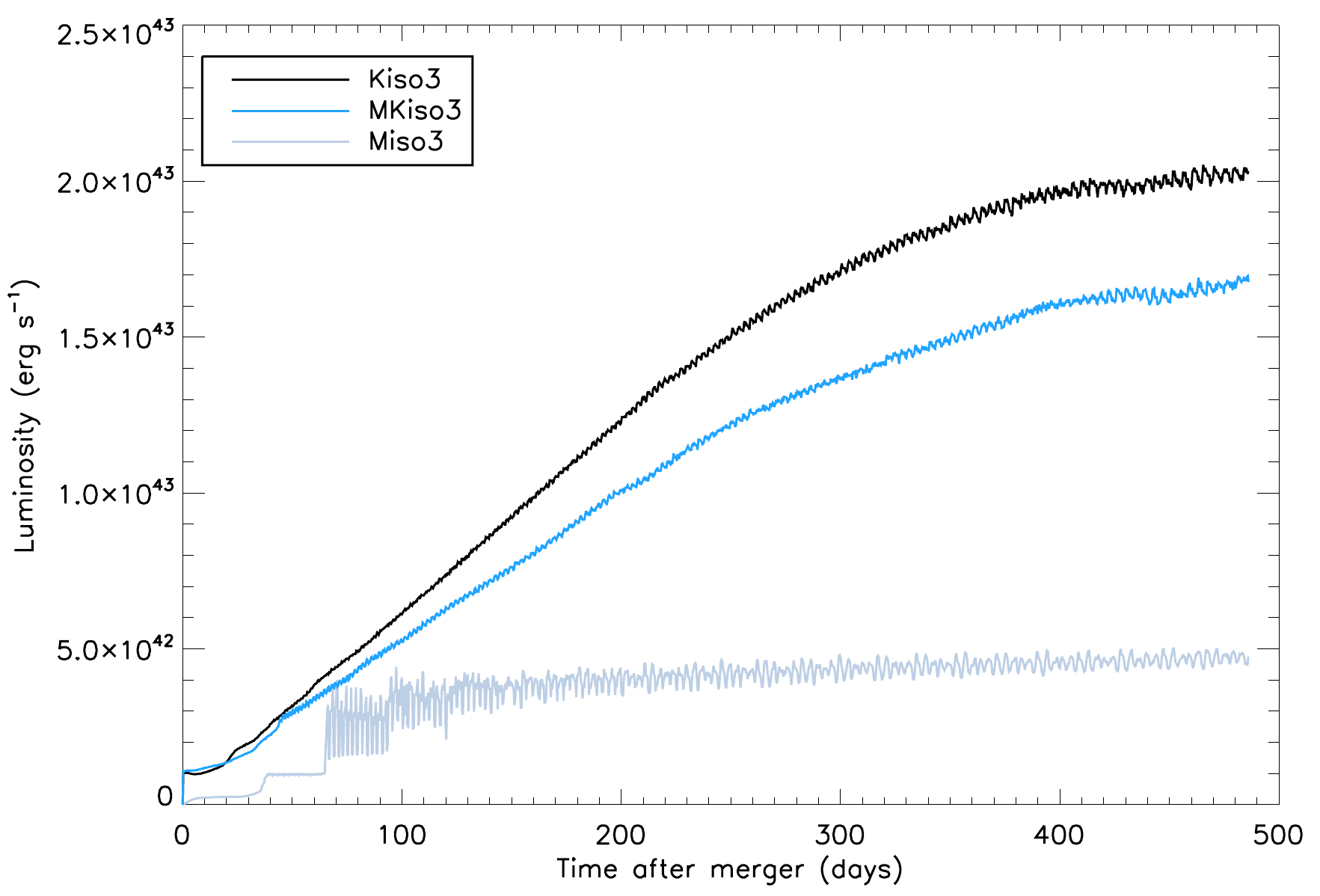}
	\caption{Light curves as in Figure~\ref{fig:LightCurves},
          except here we contrast the results from the mass-loss only
          (\textsl{Miso3}), mass-loss + kick (\textsl{MKiso3}), and
          the kick-only cases (\textsl{Kiso3}).  When the effect of
          mass-loss is included in addition to the kick, it reduces
          the overall luminosity of the disk by a modest factor.  In
          the case of mass-loss only, the luminosity of the disk is
          reduced significantly (by a factor of $\approx4$). }
	\label{fig:MKlumi}
\end{figure}

When the effects of mass-loss are included in the simulation, in
addition to the kick, the luminosity of the disk tends to decrease.
This can be expected from Figure~\ref{fig:MassLoss}, which showed that
the overdensities in the spiral shock waves tend to be reduced
compared to the case without mass-loss.  Figure~\ref{fig:MKlumi}
illustrates the effects of mass-loss on the light curve produced from
a kicked disk, as well as the luminosity produced in the mass-loss
case.

\begin{figure}
\centering
	\includegraphics [scale=0.4] {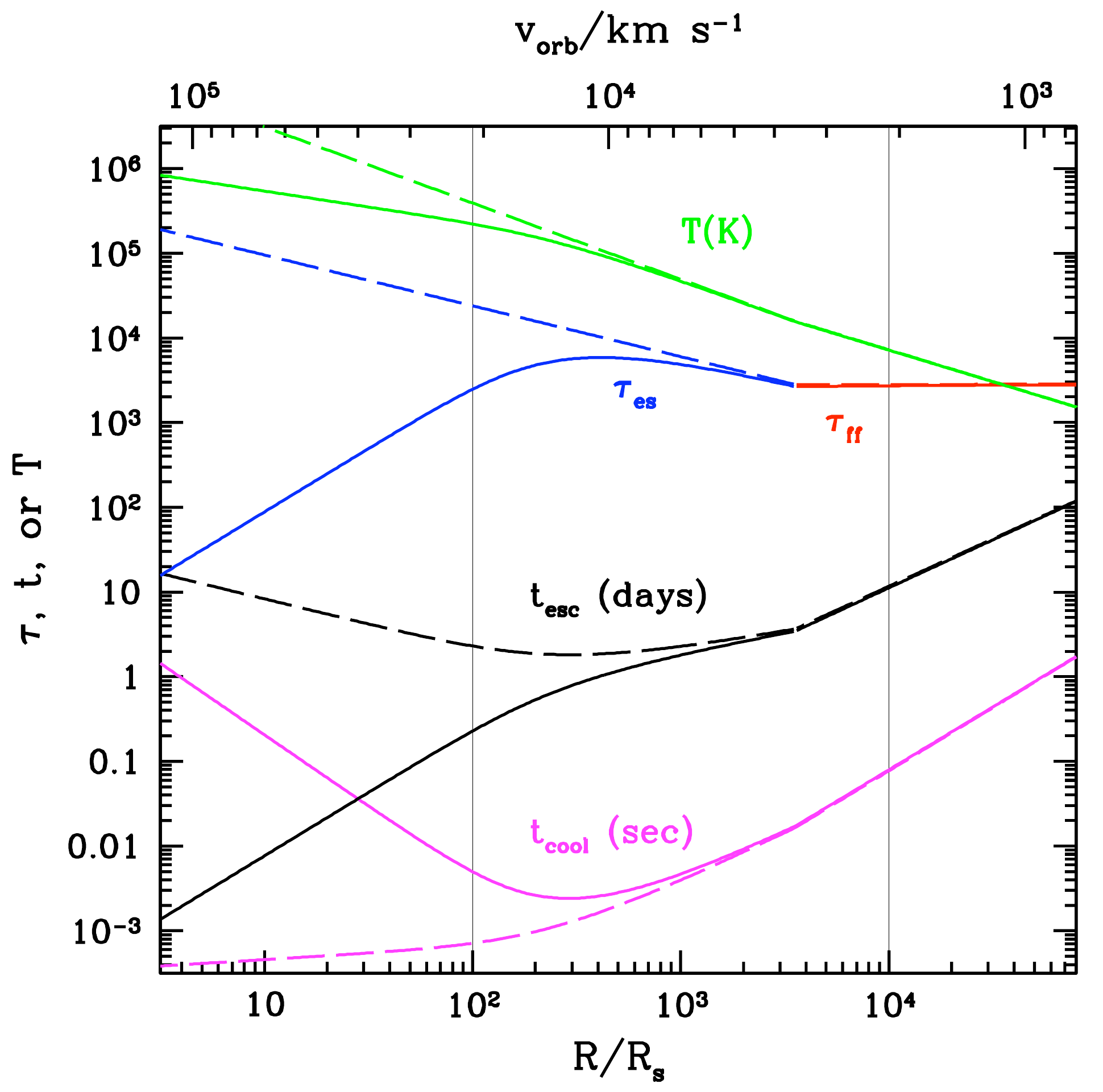}
	\caption{Physical quantities in a standard $\alpha$--disk
          around a $M=10^6 M_{\rm \odot}$ black hole, as a function of
          radius (lower labels) and corresponding orbital velocity
          (upper labels). We show the temperature ($T$; in Kelvin;
          green curves), the vertical optical depth ($\tau$; blue/red
          when electron scattering/free--free opacity dominates,
          respectively), the Bremsstrahlung cooling time ($t_{\rm
            cool}$; in units of seconds; magenta curves) and the
          vertical photon escape time ($t_{\rm esc}$; in units of
          days; black curves).  Solid curves are for a standard
          $\alpha$--disk (with viscosity scaling with total gas +
          radiation pressure), and the dashed curves are for a
          standard $\beta$--disk (with viscosity scaling with gas
          pressure.  The thin vertical lines bracket the extent of the
          high--resolution mesh in our simulations.}
        \label{fig:PhotonEscape}	
\end{figure}

The above estimates for the light--curve utilize only the effective
dissipation that occurs, implicitly, in our simulations.  The photons
generated by this dissipation must escape the disk before they can be
observed.  We next use standard $\alpha$--disk models to estimate the
basic parameters of a more realistic disk.  We adopt the equations
describing the radial profile of the density, scale--height,
temperature, and opacity of a standard thin accretion disk as
summarized in \citet[][based, in turn, primarily on Frank et al. 2002
and on Goodman \& Tan 2004]{HKM09b}.  In
Figure~\ref{fig:PhotonEscape}, we show the temperature ($T$; in
Kelvin; green curves), the vertical optical depth ($\tau$; blue/red
when electron scattering/free--free opacity dominates, respectively),
the cooling time due to Bremsstrahlung emission ($t_{\rm cool}$; taken
from eq. 6.53 in \citealt{accretionpower} in units of seconds; magenta
curves) and the vertical photon escape time ($t_{\rm esc}$; in units
of days; black curves).  The vertical optical depth is taken to be
$\tau={\rm max}(\tau_{\rm es},\tau_{\rm ff})$, where $\tau_{\rm es}$
is the free--free optical depth from Kramer's opacity, and $\tau_{\rm
  es}$ is the electron scattering optical depth. The photon escape
time is obtained as the product of the mean--free--path $H/\tau$
(where $H$ is the vertical scale--height) and the number of
scatterings in a random walk required to escape ($\tau^2$), divided by
the speed of light, $t_{\rm esc}=(H/c)\tau$.  All curves are for a
$M=10^6 M_{\rm \odot}$ black hole.  Solid curves are for a standard
$\alpha$--disk (with viscosity proportional to the total pressure of
gas + radiation), and the dashed curves are for a $\beta$--disk (with
viscosity proportional to the gas pressure).  The viscosity parameter,
the accretion rate (in Eddington units), and the radiative efficiency
were chosen to be $\alpha=0.3$, $\dot{m}=0.1\dot{M}_{\rm Edd}$ and
$\epsilon=0.1$, respectively.  All the other parameters, which have a
relatively minor effect on the profiles, are the same as the fiducial
ones defined in \citet{HKM09b}.

As Figure~\ref{fig:PhotonEscape} shows, the disk is optically thick
($\tau\gg1$) at all relevant radii, and at radii $\lsim 8000 R_s$, the
opacity is dominated by electron scattering.  Near this radius, the
local Bremsstrahlung cooling time is negligibly short, and the photon
escape time from the disk is only $\sim 3$ days.  Therefore, even
though the disk is optically thick, the photons, generated by
dissipation in the mid--plane, will emerge from the disk with a
relatively short delay.  These conclusions are derived for a steady
disk. In the case of a binary, the pile-up of material near $100 R_s$
will modify the disk structure, corresponding, locally, to a
non--accreting ($\dot{M}=0$) disk \citep{ivanov,mp05}; however, the
steady-disk model should give a good order--of magnitude estimate
farther out at $\sim 1000 R_s$, where the disk is relatively less
disturbed before the merger.

\begin{figure}
\centering
	\includegraphics [scale=0.45] {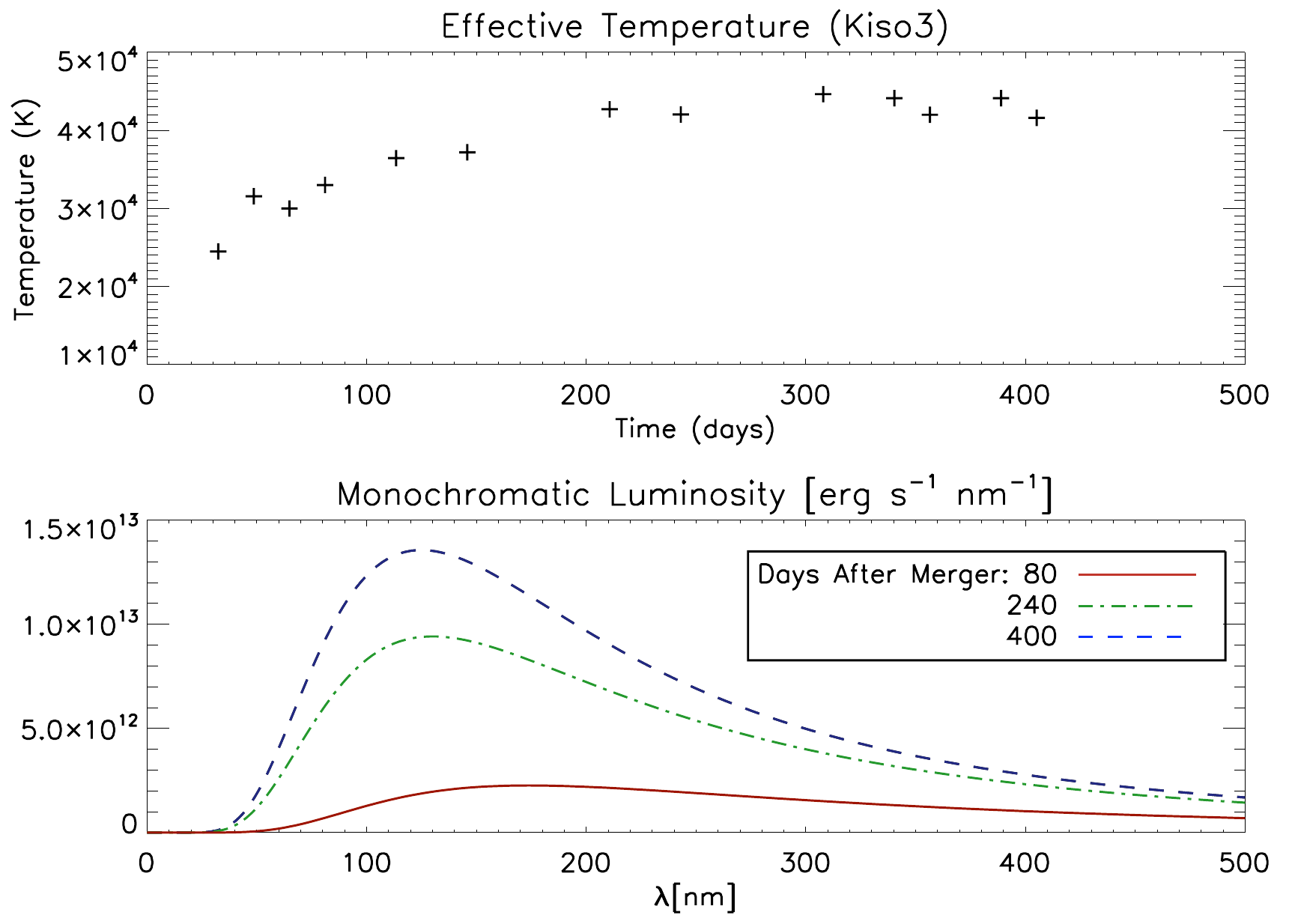}
	\caption{{\it Top panel:} effective black--body temperature
          $T_{\rm eff}$ for the most luminous point in the disk in the
          fiducial \textsl{Kiso3} run, defined by requiring the
          black--body surface brightness $\sigma T_{\rm eff}^4$ to
          equal the dissipation rate derived in the simulation. {\it
            Bottom panel:} total monochromatic luminosity emerging
          from our fiducial \textsl{Kiso3} disk at three different
          times, obtained by integrating the local black body emission
          over the face of the disk. The three curves (from bottom to
          top) show the spectra at $t=80$, 240, and 400 days after the
          merger.}
	\label{fig:Spectrum}
\end{figure}

Since the disk is very optically thick, the spectrum that emerges will
be significantly processed, and close to a black--body shape
(sometimes referred to as ``grey--body'';
e.g. \citealt{mp05}). Assuming for simplicity that the radiation
escaping from the surface has a perfect thermal spectrum, we can
compute the required effective temperature $T_{\rm eff}$, by equating
the black--body luminosity per surface area $\sigma T_{\rm eff}^4$ to
the luminosity density $j$ we derive from the simulations.  In the top
panel of Figure~\ref{fig:Spectrum}, we show the evolution of this
effective temperature, computed as a function of time at the most
luminous point in the disk.  As the figure shows, $T_{\rm eff}$ rises
and asymptotes to $\sim 4.5 \times 10^4$ K, a value that is a factor
of $\sim 3$ lower than the constant isothermal temperature $T^{\rm
  iso}$ imposed on the disk.  This is consistent, to within a factor
of $\sim$two, with the vertical structure in thin accretion disk
models, in which the surface temperature is lower by a factor of $\sim
\tau^{-1/4}$ than the mid--plane temperature
\citep[e.g.][]{accretionpower}.

From the similarity of the effective temperatures, we infer that the
transient dissipation required to maintain a constant temperature in
the disturbed disks is comparable (albeit an order of magnitude
higher) than the viscous dissipation that occurs in a standard,
steady-state, thin accretion disk.  This implies that if the
pre-merger disk luminosity was comparable to that of a thin accretion
disk, the merger would be accompanied by an order-of-magnitude rise in
luminosity.  In practice, the total luminosity of a steady disk is
dominated by its inner regions, and the pre--merger luminosity of the
circumbinary disk can be significantly lower (and correspondingly, the
merger will result in brightening by a much larger factor).

Interestingly, the monotonic rise in $T_{\rm eff}$ suggests that the
shock--induced emission may have the unique signature of {\it
  hardening} in frequency by a factor of $\sim$ two (in addition to
monotonically rising in luminosity by a factor of a few) within the
year following the merger event.  In the bottom panel of
Figure~\ref{fig:Spectrum}, we show the emergent composite spectrum,
integrating the local black body emission over the face of the disk,
which shows the same two features: the disk brightens, and the
spectrum hardens with time.  If confirmed by more accurate modeling of
the emergent spectrum, this should greatly help the identification of
EM counterparts, since other types of known transient UV/optical/IR
sources do not have these qualitative features.

\subsection{Varying the Initial Disk Profile}
\label{sec:Alphadisks}

All results in the previous sections are derived, for simplicity, for
disks with constant initial density and temperature.  In order to
quantify the sensitivity of our results to the initial profiles, we
have re-run several of our simulations with the power--law profiles in
equations~(\ref{eq:Sigslope})-(\ref{eq:Tslope}), approximating those
in a more realistic, standard $\alpha$--disk.  The spatial resolution
of all the $\alpha$--disk run is set to be the same as in the
highest--resolution constant density case, \textsl{Kiso3}$^{++}$, with
$(\Delta r,\Delta\theta)= (5~R_s, 0.0025~{\rm rad})$.
Table~\ref{tab:AlphaSims} summarizes the $\alpha$--disk runs.

\begin{figure}
\centering \includegraphics [scale=0.45] {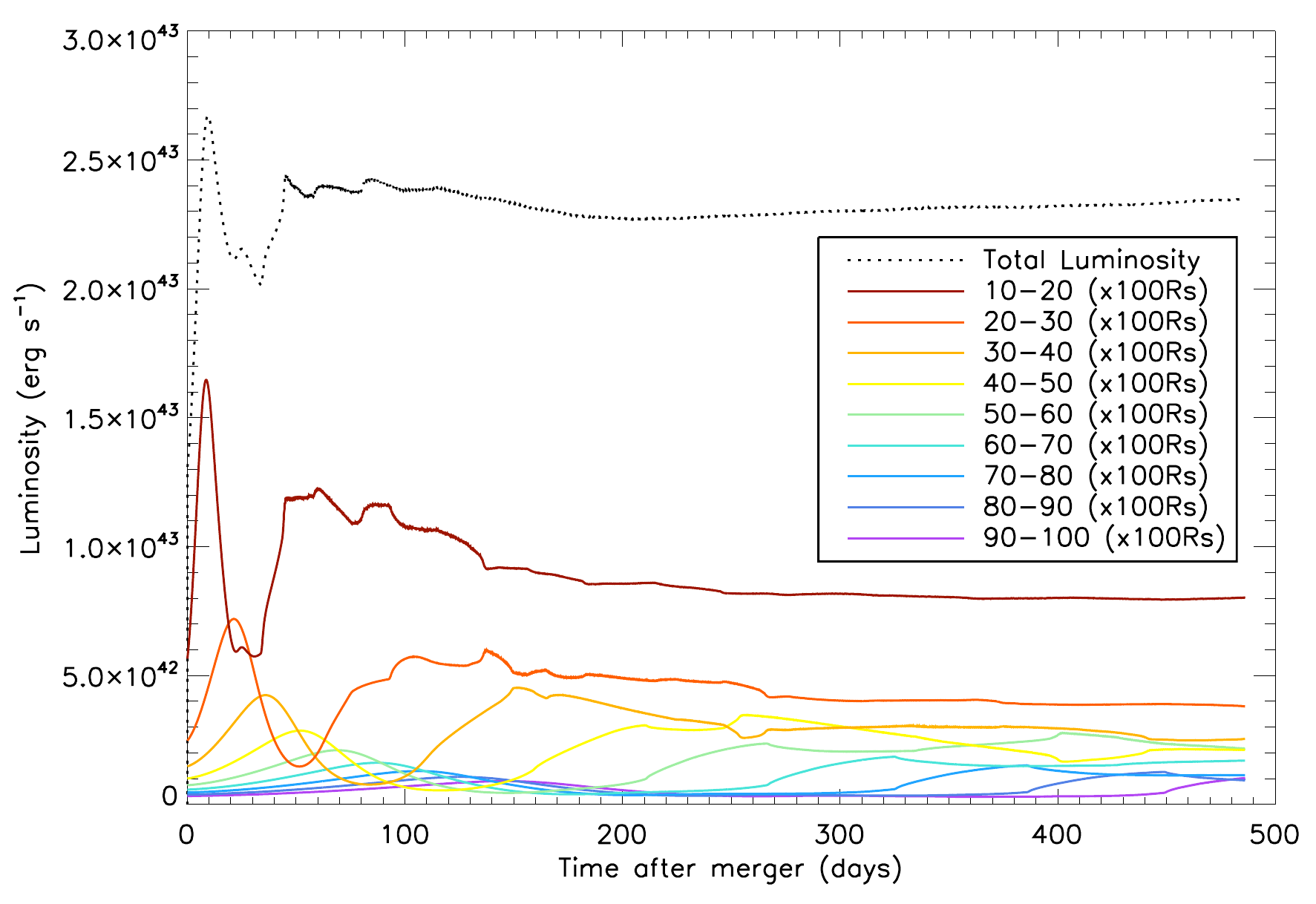}
\caption{Total luminosity (black dotted line) compared to the
  luminosity contribution from different annuli (solid colored lines)
  in an $\alpha$--disk that received a 530 km s$^{-1}$ kick and
  experienced 5\% mass loss from the central black hole
  (\textsl{MKiso}).  The initial flares occur when the first two arms
  of the shock pass through the disk.  However, the luminosity of the
  outer region of the shock decreases dramatically because less gas is
  swept up by the shocks in the outer regions of the disk, which is
  more diffuse.  A large number of thin, tightly wound density waves
  persist throughout the entire disk, and the disk maintains a nearly
  constant luminosity until the end of the simulation.}
	\label{fig:MKisoLightCurve}
\end{figure}

Figure~\ref{fig:MKisoLightCurve} shows the total luminosity of the
$\alpha$--disk, again conservatively from the restricted region $10^3
< R/R_s < 10^4$ (black dotted line) together with the luminosity
contribution from different annuli (solid colored lines).  The first
peak in each annulus, which is easily identifiable around $t = 10$
days in the $1000 < R/R_s < 2000$ bin, comes about when the first arm
of the spiral shock forms.  The second winding on the spiral arm,
which is considerably narrower at early times than the first spiral
arm, accounts for the second rise in the binned luminosity ($\sim t =
50$ days in the first bin).  As the spiral pattern progresses outward,
tighter windings of the spiral pattern persist and are generally long
lasting due to the lack of viscous dissipation in the disk.  As a
result, after the first few windings of the spiral arm pass, each
annulus contributes a steady background luminosity due to the numerous
spiral density ripples that persist to the end of the
simulation. While this is true in the uniform disk runs, as well, here
the innermost annuli continue to dominate the total luminosity at all
times, even though the outer annuli cover a larger surface area of the
disk.  It is also noticeable that, as the first few spiral arms
propagate outward, their contribution contributes less and less to the
total luminosity (compare Figure~\ref{fig:MKisoLightCurve} to
~\ref{fig:BinLumi}, where the flat portion of the
lightcurve--contribution from each bin is due to the densest point
moving into the respective region.)  In the $\alpha$--disk simulation,
the front portion of the spiral shock contributes less to the total
luminosity as it moves outward because the surface density of the gas
is considerably lower at larger radii.

\begin{figure}
\centering \includegraphics [scale=0.45] {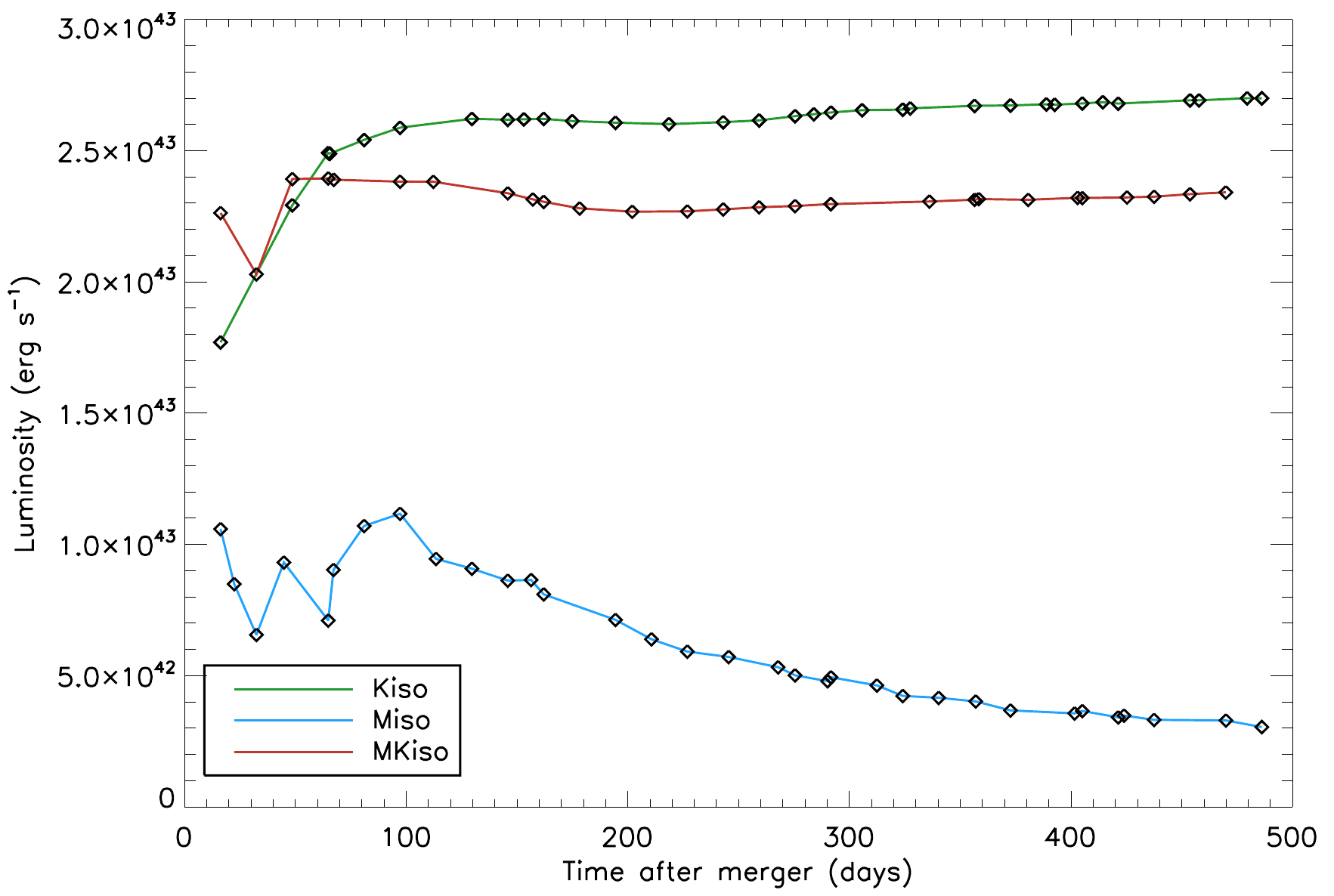}
\caption{Light curves for three isothermal disks, experiencing kick
  and no mass--loss (\textsl{Kiso}), mass--loss and no kick
  (\textsl{Miso}), and both kick and mass--loss (\textsl{MKiso}) as in
  Figure~\ref{fig:MKlumi}, except the disks here have initial
  power--law density and temperature profiles, appropriate for a
  standard $\alpha$--disk, rather than constant values.  The
  luminosities rise more rapidly, due to the higher initial density
  concentration near the central regions, but settle at values similar
  to those reached in the constant density and temperature case.}
	\label{fig:AlphaLightCurves}
\end{figure}

Figure~\ref{fig:AlphaLightCurves} shows the light--curves in the three
isothermal $\alpha$--disk simulations.  The absolute value of the
luminosities arising from shocks in the $\alpha$--disks is very close
to the previous results in the constant density disk cases.  However,
there are a few qualitative differences worth noting.  First, the
luminosity rises more rapidly in the $\alpha$--disks, reaching peak
brightness in $\approx$ 100 days, compared to the constant--density
disks, in which the peak is reached only after $\approx$ 300 days.
While some of this more rapid rise can be attributed to the higher
resolution employed for the $\alpha$--disks, the resolution study in
Figure~\ref{fig:LightCurves} makes it clear that most of the
difference is caused by the change in the profiles.  This qualitative
difference is unsurprising, since the $\alpha$--disks are more
centrally concentrated, and have more material in the inner regions,
increasing the luminosities at early times.  The mass--loss only case
shows that a brighter and more prompt flare erupts after the merger,
compared to the constant density disk; again this is attributable to
the larger amount of material near the central regions.  When both
mass--loss and a kick is included, we find a prompt flare, followed by
a nearly steady luminosity, reduced by around $15\%$ from the
kick--only case.

Overall, the light--curves in the $\alpha$--disk and constant--density
cases are reassuringly similar, indicating that variations in the
assumed initial disk profiles do not lead to order--of--magnitude
changes in the predicted brightness of the after--glow.

\subsection{Numerical Issues}
\label{sec:Numerical}

In this section, we discuss numerical effects from the inner and outer
boundary of the simulation box, the dependence of our results on the
spatial and time resolution, as well as numerically scaling our
results to apply to black holes and disks with different physical
sizes and kicks with different speeds.

{\it Boundary conditions.}  Both at the inner and the outer
boundaries, we imposed ``outflow'' boundary conditions that maintain a
zero pressure and density gradient at the edges of the simulation.  To
test the effect of these boundary conditions, we checked whether an
unperturbed disk (without any kick or mass--loss) maintains its
uniform circular Kepler motion.  We found that the outer disk, beyond
$R_s = 10^3 R_s$, indeed remained undisturbed, but modest density
ripples (with over-densities $\Sigma/\Sigma_i \lsim 2$) were produced
artificially inside $10^3 R_s$.  The amplitude of these numerical
fluctuations are below the density enhancements excited by shocks in
the disk (see for example Figures~\ref{fig:DensRad} and
\ref{fig:Rshock}).  Nevertheless, this low--amplitude noise covers a
large fraction of the surface area of the inner disk, and can dominate
the total luminosity in the inner regions.  This is a concern
especially for the $\alpha$ disks, in which the inner regions were
also found to dominate the total luminosity. More specifically, we
computed the luminosity inside $R_s = 10^3 R_s$ in the unperturbed
constant density and $\alpha$--disk models, and we found that they
reach values larger than (in the constant surface density disk cases)
or comparable (in the $\alpha$--disk cases) to the luminosities
exhibited in the perturbed disks.  In future work, we will attempt to
identify the origin of this numerical noise, and to reduce its
contribution to the luminosity.  In the present paper, we simply
derived luminosities outside $R > 10^3 R_s$, which are safe from
numerical noise, but represent an underestimate the true total
luminosity.

A separate concern is that due to the finite size of our simulation,
the outermost regions of the spiral shock wave eventually reach the
outer boundary of our region of interest. In the \textsl{Kiso3} run,
the outer edge of the high--resolution mesh is reached at $t\approx
250$ days.  Although the spiral shocks are subsequently still present
farther out in the low--resolution mesh, we have excluded these poorly
resolved shocks from our analysis and results.  For example, the dense
spiral arms moving out of the simulation box produce the turn--over in
the percentage of the disk undergoing a shock, shown in
Figure~\ref{fig:PercentShock} (the figure only includes shocked gas
within $R < 10^4 R_s$).

{\it Numerical resolution.}  In order to test the numerical
convergence of our results, we re--computed the light--curves in our
fiducial disk (\textsl{Kiso3}) at a factor of two lower
(\textsl{Kiso3}$^-$) and factors of two and four higher
(\textsl{Kiso3}$^+$ \& \textsl{Kiso3}$^{++}$) spatial
resolutions. Note that FLASH chooses time steps adaptively, matched to
the spatial resolution, so each resolution test differs from the next
by a factor of two in the time step, as well.  The four curves in
Figure~\ref{fig:LightCurves} show the light--curves in each of these
four runs, corresponding, from bottom to top, to increasing
resolution.  The light--curves begin to converge, in the sense that
their fractional difference between pairs of resolutions begins to
decrease. There is still a noticeable difference (by a relatively
small fraction of $\approx 10\%$) between the luminosities computed in
our highest and next--to--highest resolution run.  Given that the
total luminosity is dominated by sharp features, it is not surprising
that the disks are more luminous in the higher resolution runs -- the
density spikes are better resolved in these runs.  At our fiducial
resolution, the luminosities are underestimated by $\gsim 35\%$.
Unfortunately, the computational time required to run the highest
resolution simulations (which took approximately 2 weeks on 80
processors) was prohibitively expensive, and prevented us from further
increasing our resolution, in order to achieve full convergence.
However, the 10\% difference between the two highest resolution runs
is modest in comparison to other uncertainties arising from the highly
idealized nature of our luminosity estimates. The resolution study
also shows that our somewhat under--resolved simulations are
conservative, underestimating the true luminosity.

\begin{table*}
  \caption{\it Conversion from the physical parameters in our fiducial
    run to other values, for each parameter specified in our initial
    conditions, as well as for the luminosity and energy derived from the
    simulations. }
\label{tab:conversion}
\begin{center}
\begin{tabular} {l c c c}
	\hline
	Simulation Parameter & Conversion Factor ($fX^{\rm sim}$) & $\gamma_m$ & $\gamma_v$ \\
	\hline
	BH Mass (${\rm M_\odot}$) & $10^{6} $ & 1 & 0 \\
	Radius of inner disk edge (cm) & $ 3.0 \times 10^{13} $ & 1 & -2 \\
	Maximum time (days) & $500$ & 1 & -3 \\
	Kick velocity (${\rm km~s^{-1}}$) & $530 $ & 0 & 1 \\
	Gas temperature (K) & $1.4 \times 10^{5} $ & 0 & 2 \\
	Surface Density (g cm$^{-2}$) & $1.5 \times 10^{5} $ & -1 & 4 \\
	\hline
	Luminosity (erg s$^{-1}$) &  -- & 0 & 5 \\
	Energy (erg) &  -- & 1 & 2 \\
	\hline
\end{tabular}
\end{center}
\end{table*}

{\it Scaling simulation runs to physical units.}  The physical scales
in each simulation are fixed by specifying the physical values of any
two independent dimensional parameters -- these can be taken to be the
BH mass $M$, and the kick velocity $v_{\rm kick}$.  Our runs can
describe systems with different combinations of BH masses and kick
velocities, provided the other parameters are suitably adjusted.
Consider the quantity $X$, specified in the simulations by a
dimensionless value $X^{\rm sim}$. The physical value $X^{\rm phys}$
of this quantity can be obtained by a constant conversion factor $f$,
and will scale with $M$ and $v_{\rm kick}$ as
\begin{equation}
	\frac{X^{\rm phys}}{X^{\rm sim}} = f \left( \frac{M}{10^6 M_\odot} \right)^{\gamma_m} \left( \frac{v_k}{530 \mbox{ km/s}} \right) ^{\gamma_v}.
\end{equation}
Table \ref{tab:conversion} shows the factor $fX^{\rm sim}$ for various
parameters specified in the initial condition of our fiducial run, and
the mass-- and velocity--dependence of these parameters; we also show
the mass-- and velocity--dependence of the physical luminosity and
energy, derived from the simulations.  For example, as $v_{\rm kick}$
increases, $v_{\rm orb}$ must increase in proportion, decreasing both
the distance ($\propto v_{\rm kick}^{-2}$) and time ($\propto v_{\rm
  kick}^{-3}$) scales.  Thus, our results would be applicable, for
example, for $M=10^6~{\rm M_\odot}$ and $v_{\rm kick}=10^3~{\rm
  km~s^{-1}}$, but would then describe the evolution of a 4 times
smaller disk, with an inner cavity of size $25 R_s$, and for only
$\approx 50$ days, rather than a year. The disks would also be 4 times
hotter, and 16 times more massive than for the fiducial 530 ${\rm
  km~s^{-1}}$ kick.  The luminosity has the especially steep
dependence $\propto v_{\rm kick}^5$, so that higher--velocity kicks
would be much more observable (provided, of course, that the scaled
physical parameters are realized in nature).  Other combinations to
which our runs can correspond can be read off similarly from Table
\ref{tab:conversion}.

\subsection{Comparison with Previous Work}
\label{sec:Others}

\citet{Lippai} considered the response of a collisionless disk to
kicks.  Qualitatively, the spiral shock structures shown in
Figure~\ref{fig:diskdens} in our fiducial \textsl{Kiso3} run are very
similar to the caustics shown in Figure~1 of \citet{Lippai} for a
system with the same parameters (the only difference is that
\citealt{Lippai} used $v_{\rm kick}=500~{\rm km~s^{-1}}$).  However, a
more quantitative comparison is useful to separate hydrodynamical
effects from those arising from orbital dynamics.

Figure~\ref{fig:Rshock} shows the radial position of the densest point
in the simulated disks, compared to the location $r_c$ of the
outermost caustic identified in a collisionless case
\citep{Lippai}. The straight dotted lines correspond to propagation at
constant speeds of $v_{\rm kick}=530~{\rm km~s^{-1}}$ (lower line) and
$2v_{\rm kick}=1060~{\rm km~s^{-1}}$ (upper line).  As the figure
demonstrates, the shocks found in our isothermal runs form at
locations very close to those of the collisionless caustics. Both
follow the approximate simple linear expansion $r_c\approx v_{\rm
  kick}t$ expected from the epicyclic approximation \citep{Lippai},
although the propagation is somewhat faster at early times, and
somewhat slower at late times. Apparently, while the temperature has a
strong impact on the overdensities and the strengths of the shocks
(shown, e.g., in Fig.~\ref{fig:DensRad}) it makes a comparatively
small difference to their propagation speed or to their overall
morphology.  These results also provide reassurance that the shock
propagation was not affected by numerical edge effects from the inner
and outer boundary of the simulation.  Interestingly, on the other
hand, the equation of state has a significant impact, with the
propagation $\sim$ twice as rapid in the adiabatic runs.  As mentioned
above, this apparent difference in the propagation speed of the
densest point arises partly because the densest point falls further
out along the outermost spiral arm in the adiabatic runs, and partly
because the gas pressure makes the outermost spiral arms more diffuse
and extend to a larger radius.

As mentioned in \S~\ref{sec:PostMerger} above, we found that pressure
suppresses density enhancements in the disk at late times, by large
factors.  This last conclusion is somewhat surprising, given
expectations from collisionless disks. \citet{Lippai} found that the
relative velocity $v_c$ of particles, as they cross on their orbits to
create the caustics, increases with increasing radius as $v_c\sim
v_{\rm kick}^2/v_{\rm orbit}\propto r^{1/2}$ (where $v_{\rm orbit}$ is
the orbital velocity). The collisions therefore become super--sonic
beyond some critical radius; this critical radius typically lies
within the range of radii shown in Figure~\ref{fig:DensRad}. For
example, in the fiducial run \textsl{Kiso3}, the critical radius is at
$1,300 R_s$, and in the higher--temperature runs \textsl{Kiso4} and
\textsl{Kiso5}, it is at $4,200 R_s$ and $13,000 R_s$, respectively.
We find that the large overdensity of the shocks can be attributed to
``pre--compression''. In particular, the shocks exceed the overdensity
of a factor of 4 relative to the surface density at the start of the
simulations. This is because significant overdensities start building
up along the spiral pattern before the actual shocks occur.  This
increases the absolute density of the shock, since the shock is
hitting gas that already has a surface density higher than the initial
value.  For example, in Figure~\ref{fig:diskdens}, the adiabatic
simulation (\textsl{Kad3}) shows that the outermost edge of the spiral
pattern has overdensities forming with $\Delta\Sigma \approx 2$.  For
a strong shock and $\gamma=5/3$, we expect a density enhancement about
4 times that of the initial density, or $\Sigma \approx 8\Sigma_i$,
which is what we see in the denser part of the spiral.  A similar
mechanism is at work in the isothermal simulation (\textsl{Kiso3}),
but the shocked gas is able to cool and reach much larger densities.
This pre--compression, which happens on time--scales much longer than
the shock--formation itself, can apparently be significantly
suppressed by gas pressure.

\citet{Lippai} also proposed that the shock temperature scales with
time/radius as $T_{\rm shock}\propto v_{\rm c}^2 \propto v_{\rm
  orbit}^{-2} \propto r_c \propto t_c$, and suggested that the
luminosity is dominated by the outermost shocked shells, with both the
total luminosity and the characteristic frequency of the emitted
spectrum increasing with time. The same qualitative behavior is seen
in our isothermal, constant--density simulations.  However, we find
that the largest Mach number in the spiral arms increases from
$\mathcal{M}=1.7$ to $\mathcal{M}=3$, between $t=16$ and $310$ days,
much less steeply than linearly with time (and radius).  In our
adiabatic runs, the Mach numbers decline from $2$ to $1.1$ in the same
interval. This decline, however, is due to the increase in sound
speed, rather than a decrease in the gas velocities.

Assuming a disk profile with $\Sigma\propto r^{-3/5}$ (close to the
one adopted in our $\alpha$--disk runs), and also that a constant
fraction of the disk mass interior to the outermost spiral caustic is
shocked ($M_{\rm shock} \propto \Sigma r dr$) and that the energy is
released over the time--scale of $t_c=r_c/v_{\rm kick}$,
\citet{Lippai} also finds the scaling $L_{\rm kick} \approx (1/2)
M_{\rm shock} v_c^2 / t_c \propto r^{19/10}$.  As discussed in
\S~\ref{sec:Observational} above, the disk is likely to be optically
thick, in which case the spectrum is significantly re--processed,
invalidating naive conclusions about the optically--thin spectrum.  In
the constant--density runs, we nevertheless find an approximately
linearly rising luminosity, and also a modestly increasing
characteristic frequency.  While this is in qualitative agreement with
the above expectations, most of the rise in the luminosity in our case
is explained by the increasing fraction of the disk mass that is
shocked, with the increase in the characteristic temperature
contributing relatively little.  In our $\alpha$--disk runs, we find a
light--curve that initially rises and then reaches a nearly constant
luminosity after $\approx$100 days.

As mentioned in the Introduction, \citet{ONeill} and \citet{Megevand}
have recently employed three--dimensional simulations to study the
evolution of much smaller disks (extending from $2-50 R_s$ and $10-30
R_s$, respectively, compared to our $10^2-10^4 R_s$) on much shorter
time--scales ($\sim 5$ hrs and $\sim 6$ hrs, scaling to a $10^6~{\rm
  M_\odot}$ BH; compared to our $\sim 500$ days).  Since our FLASH
simulation results are output at times that sample the disk evolution
only every $\approx 4$days, we can not directly compare the behavior
of our disks to those in \citet{ONeill} and \citet{Megevand}.

Despite the mis--match in scales and different dimensionality, it is
interesting to contrast the light--curves we find here with those in
\citet{ONeill} and \citet{Megevand}.  We first note that the
simulations in both \citet{ONeill} and \citet{Megevand} are adiabatic,
i.e., they are energy--conserving by construction.  \citet{ONeill} and
\citet{Megevand} present estimates for the luminosity by computing the
volume--integral of the quantity $\rho^2 T^{1/2}$, which is
proportional to the local Bremsstrahlung emissivity.  As we discussed
in \S~\ref{sec:Observational} above, the Bremsstrahlung cooling time
is exceedingly short, so if dissipation indeed tracked the
Bremsstrahlung emission, the adiabatic condition could not be
maintained.  If cooling were efficient, then one expects, instead,
much stronger overdensities to develop than in the adiabatic case,
which would, in turn, generically increase the cooling rate
$\propto\int dV \rho^2 T^{1/2}$.

Setting these inconsistency issues aside, in order to compare our
results as directly as possible to \citet{ONeill} and
\citet{Megevand}, we computed the quantity $L_{\rm Brem}=\int dV
\rho^2 T^{1/2}$ in both our adiabatic and isothermal simulations.
Unfortunately, this comparison is further complicated by the fact that
our simulations are two--dimensional.  In order to calculate the 3D
gas densities and volume elements, $\rho$ and $dV$ from the constant
density disks, we assumed that the disk has the scale height $H = c_s/
\Omega$ obtained by assuming hydrostatic equilibrium in the vertical
direction (here $c_s = \sqrt{k T/\mu m_p}$ and $\Omega = v_\phi/r$ are
measured directly in the simulation run). This assumption should be
reasonable, since we simulated $N\gg 1$ dynamical times, except near
the outer edge of the disk.  Note that if hydrostatic equilibrium was
not established (i.e. the disk was still expanding in the vertical
direction, following the mass--loss or recoil), the disk would be
thinner, and the 3D densities would be even higher than our estimates,
increasing the predicted luminosities.  Finally, note that under the
assumption of hydrostatic equilibrium, and with $\rho=\Sigma/H$, the
Bremsstrahlung integral becomes independent of temperature, $\int dV
\rho^2 T^{1/2}=$ const $\int dr \Sigma^2 r^{-1/2}$.

\begin{figure}
\centering
	\includegraphics [scale=0.5] {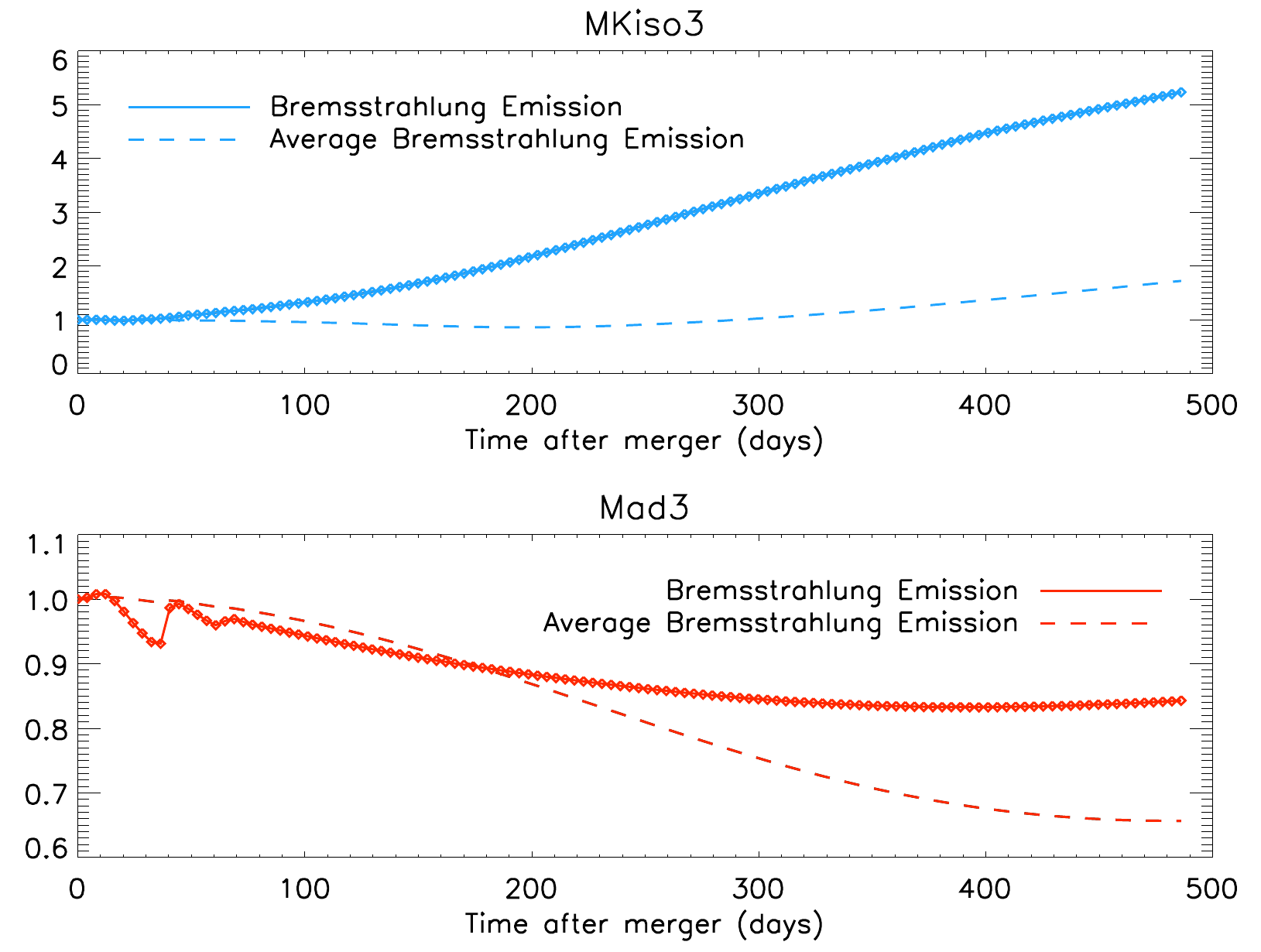}
	\caption{{\em Lower panel:} Thermal Bremsstrahlung emission
          from an adiabatic, constant--density disk that experiences
          the effect of 5\% mass loss and no kick (\textsl{Mad3}). The
          luminosity is normalized to its steady pre--merger value.
          Initially, the light curve exhibits variability,
          qualitatively similar to the dips and troughs found, on
          shorter time--scales, by both \citet{ONeill} and
          \citet{Megevand}.  The dashed curve shows, for comparison,
          the Bremsstrahlung luminosity that would result from a disk
          with uniform surface density and the same total mass.  The
          prominent dip at $t>100$ days is caused by $\approx$20\% of
          the original disk material exiting the simulation volume.
          {\it Upper panel:} Bremsstrahlung emission from our fiducial
          isothermal disk, experiencing both mass--loss and a kick
          (\textsl{MKiso3}).  The dashed curve shows the corresponding
          luminosity from a uniform disk, as in the lower panel. The
          disk brightens, mostly due to the sharp density contrasts
          that develop.}
	\label{fig:BremCDBoth}
\end{figure}

The solid curve in bottom panel in Figure~\ref{fig:BremCDBoth} shows
the evolution of $L_{\rm Brem}$ in the mass--loss only, constant
density disk runs.  The light curve exhibits oscillations within the
first 100 days after the merger, qualitatively similar to the dips and
troughs found, on shorter time--scales, by both \citet{ONeill} and
\citet{Megevand}. In particular, as in the those two studies, the
entire disk remains dimmer than prior to the mass--loss: the
luminosity of the disk decays by $\approx 20\%$ over the course of a
year.  To understand this result better, we computed the
time-dependent mean surface density $\langle \Sigma\rangle$ of the
simulated disk between $10^3-10^4R_s$.  As noted above, the disk as a
whole expands after the mass-loss, due to the sudden decrease in
gravitational force from the central point mass. We show, by the
dashed curve in the bottom panel of Figure~\ref{fig:BremCDBoth}, the
Bremsstrahlung luminosity that would result from a disk with the
time--dependent surface density $\langle \Sigma\rangle_t$.  The departure of
the dashed curve from unity is therefore due entirely to changes in
the total disk mass within the region $10^3-10^4R_s$, whereas the
difference between the solid and dashed curve is caused entirely by
inhomogeneities in the gas surface density in this region.  As these
two curves show, at early times, the drop in the luminosity is due
mostly to inhomogeneities, whereas at late times, the dimming is
caused by approximately 20\% of the disk mass exiting the simulated
region.  Note that at late times, the inhomogeneities actually cause
an increase in the luminosity. At early times, inhomogeneities cause
the luminosity to decrease. While naively one expects that
inhomogeneities always boost the luminosity, we note that in the
adiabatic case, the disk is effectively allowed to expand in the
vertical direction, and inhomogeneities can cause a small overall
decrease in the luminosity.

In comparison, the top panel in Figure~\ref{fig:BremCDBoth} shows
$L_{\rm Brem}$ in our fiducial isothermal run (\textsl{Kiso3}), which
includes both a kick and a mass--loss.  As in the bottom panel, the
dashed curve shows the luminosity from a uniform disk with the same
total mass.  As this figure shows, the luminosity significantly {\em
  increases} with time, and this increase is mostly due to the
inhomogeneities (note that the density contrasts are much higher than
in the adiabatic case, and the isothermal disk does not expand in the
vertical direction). Interestingly, in this run that includes the
kick, the total mass within the $10^3-10^4R_s$ region increases, rather
than decreases.

In summary, our constant--density, adiabatic disks, which include only
mass--loss and no kick, show the same qualitative behavior as in
\citet{ONeill} and \citet{Megevand}, namely oscillations in the
luminosity, and a modest overall dimming of the disk, following the BH
merger.  However, using an isothermal, rather than an adiabatic
equation of state and considering the impact of a kick, in addition to
the mass--loss, tend to counter this dimming, and when combined,
produce, instead a significant brightening of the disk - due mostly to
the stronger density contrasts that develop in the isothermal case.

\section{Conclusions and Future Directions}
\label{sec:Conclude}

In this paper, we used two--dimensional hydrodynamical simulations of
a circumbinary disk, to follow the effects of a velocity recoil and
mass-loss of the central black hole binary, following the merger of
the two black holes. From a suite of runs, we are able to draw two
basic conclusions.

First, the outward--propagating spiral shocks that develop in our
simulations follow a pattern very similar to the caustics identified
in collisionless disk \citep{Lippai}.  Gas pressure has a modest
overall impact on the propagation and on the overall morphology of the
shocks.  On the other hand, we find that the gas pressure, and the
assumed equation of state, has a strong impact on the overdensities
that develop and on the strengths of the shocks: isothermal,
low--temperature disks have larger overdensities and stronger shocks.

Second, we have estimated an upper limit on the luminosity emerging
from the disk experiencing a BH kick with $v_{\rm kick}=530~{\rm
  km~s^{-1}}$, by measuring the effective dissipation that occurs,
implicitly, in the simulations when an isothermal condition is imposed
on the gas.  The resulting luminosity is of order $10\%$ of the
Eddington limit for the $10^6 M_\odot$ BH used in the simulation,
which suggests that the after--glow may be bright enough to be
detectable.  If the pre-merger disk has a luminosity below that of a
standard steady-state thin accretion disk (due to the evacuation of
the inner regions of the disk), then the merger-induced kick will
cause a significant (order-of-magnitude, or larger) brightening of the
disk.  We also estimated the effective black--body temperature of the
radiation emerging from the optically thick disk. We found that as the
disk brightens, the characteristic frequency increases with time,
possibly offering a unique signature of the kick--induced emission.
When the accompanying mass--loss of the merger remnant is included in
our simulations, the density contrasts and luminosity from the spiral
shocks decrease somewhat, but do not change dramatically in overall
behavior.

While our results are encouraging, and suggest that the EM signature
of the disturbed circumbinary disk may be detectable, this conclusion
has to be verified in the future in improved models. In particular, a
better estimate of the thermodynamics of the disk, with a realistic
treatment of radiative cooling, should reveal how close the luminosity
is to the values we obtained here, using the isothermal assumption; by
using three--dimensional simulations that resolve the vertical disk
structure, and by following the vertical transfer of the radiation
produced should clarify the robustness of our conclusions about the
evolution of the characteristic emergent frequency.

\section{Acknowledgments}

We are grateful to Kristen Menou, Mordecai MacLow, Greg Bryan, and
Bence Kocsis for useful discussions.  The software used in this work
was in part developed by the DOE-supported ASCI/Alliances Center for
Astrophysical Thermonuclear Flashes at the University of Chicago.
This work was supported by the Pol\'anyi Program of the Hungarian
National Office for Research and Technology (NKTH) and by the NASA
ATFP grant NNX08AH35G.

\newpage

\end{document}